%% file: main.tex
\documentclass[journal]{IEEEtran}
\IEEEoverridecommandlockouts
\usepackage{graphicx} % For including images
\usepackage{amsmath} % For math equations
\usepackage{amssymb} % For math symbols
\usepackage{cite} % For managing citations
\usepackage{hyperref} % For hyperlinks
\usepackage{bm}
\usepackage{graphicx}
\usepackage{color}
\usepackage{balance}
\usepackage{microtype}
\usepackage{acronym}
\usepackage{algorithm}
\usepackage{algpseudocode}
\usepackage{subfig}

\usepackage[normalem]{ulem}
\usepackage{tikz}
\usetikzlibrary{arrows.meta, positioning, calc, shapes, decorations.text}

\graphicspath{ {./images/} }

\newcommand{\norm}[1]{\left\lVert#1\right\rVert}

\newcommand{\diag}[1]{{\mathrm{diag}}\left(#1\right)}

\allowdisplaybreaks

\input{acronyms}

\usepackage[font = footnotesize]{caption}
\begin{document}
% Title and author information
%\title{Fundamental Performance Bounds for  multi-band Carrier Phase-based Positioning}
%\title{Fundamental Bounds for Multi-band Carrier Phase-based Positioning toward 6G Cellular}
%\title{Multi-band Carrier Phase-based Positioning: Performance Bounds and Design Insight Towards 6G Cellular}
%\title{Multi-band Carrier Phase Positioning: Performance Bounds and Design Insight Towards 6G Cellular}
\title{Multi-band Carrier Phase Positioning toward 6G: \\ Performance Bounds and Efficient Estimators}
%\title{Multi-band Carrier Phase Positioning:\\Bounds and Design Insight Towards 6G Cellular}
%\author{\IEEEauthorblockN{Ehsan Shourezari\IEEEauthorrefmark{1}, Ossi Kaltiokallio\IEEEauthorrefmark{1}, Mehmet C. Ilter\IEEEauthorrefmark{1}, Jukka Talvitie\IEEEauthorrefmark{1}, Gonzalo Seco-Granados\IEEEauthorrefmark{2}, \\ Henk Wymeersch\IEEEauthorrefmark{3}, and Mikko Valkama\IEEEauthorrefmark{1}}  
%\IEEEauthorblockA{
% \IEEEauthorrefmark{1}Department of Electrical Engineering, Tampere University, Finland\\
% \IEEEauthorrefmark{2}Universitat Autonoma de Barcelona, Barcelona, Spain\\
% \IEEEauthorrefmark{3}Department of Electrical Engineering, Chalmers University of Technology, Sweden}
%Email: \{ehsanollah.shourezari,\,mehmet.ilter,\,ossi.kaltiokallio,\,mikko.valkama\}@tuni.fi, Gonzalo.Seco@uab.cat, henkw@chalmers.se }
%\thanks{This work has been supported by Business Finland under the 6G-ISAC project, by the SNS JU project 6G-DISAC under the EU's Horizon Europe research and innovation Program under Grant Agreement No 101139130, and by the Research Council of Finland under the grant \#359095.}
%}

\author{Ehsan Shourezari,
        Ossi Kaltiokallio,~\IEEEmembership{Member,~IEEE,}
        Mehmet C. Ilter,~\IEEEmembership{Senior Member,~IEEE,}
        \\Jukka Talvitie,~\IEEEmembership{Member,~IEEE,}
        Gonzalo Seco-Granados,~\IEEEmembership{Fellow,~IEEE,}
        Henk Wymeersch,~\IEEEmembership{Fellow,~IEEE},
        \\and Mikko Valkama,~\IEEEmembership{Fellow,~IEEE} 
        \thanks{Ehsan Shourezari, Ossi Kaltiokallio, Mehmet C. Ilter, Jukka Talvitie, and Mikko Valkama are with Tampere University, Finland. % (email: \href{ehsanollah.shourezari@tuni.fi}{ehsanollah.shourezari@tuni.fi}; 
        %\href{ossi.kaltiokallio@tuni.fi}{ossi.kaltiokallio@tuni.fi}; 
        %\href{mehmet.ilter@tuni.fi}{mehmet.ilter@tuni.fi}; \href{jukka.talvitie@tuni.fi}{jukka.talvitie@tuni.fi};
        %\href{mikko.valkama@tuni.fi}{mikko.valkama@tuni.fi})
        }
        \thanks{Gonzalo Seco-Granados is with Universitat Autonoma de Barcelona, Spain. 
        %(e-mail: \href{gonzalo.seco@uab.cat}{ gonzalo.seco@uab.cat})
        }
        \thanks{Henk Wymeersch is with Chalmers University of Technology, Sweden. % (e-mail: \href{henkw@chalmers.se}{henkw@chalmers.se})
        } \vspace{-3mm}
        }

\maketitle
\bstctlcite{IEEEexample:BSTcontrol}
\begin{abstract}
%Carrier phase positioning (CPP) is widely used in satellite system applications, enabling centimeter-level localization accuracy. Recently, 
In addition to satellite systems, carrier phase positioning (CPP) is gaining attraction also in terrestrial mobile networks, particularly in 5G New Radio (NR) evolution toward 6G. One key challenge is to resolve the so-called integer ambiguity problem, as the carrier phase provides only relative position information. This work introduces and studies a multi-band CPP scenario with intra- and inter-band carrier aggregation (CA) opportunities across FR1, mmWave-FR2, and emerging 6G FR3 bands. Specifically, we derive multi-band CPP performance bounds, showcasing the superiority of multi-band CPP for high-precision localization in current and future mobile networks, while noting also practical imperfections such as clock offsets between the user equipment (UE) and the network as well as mutual clock imperfections between the network nodes. A wide collection of numerical results is provided, covering the impacts of the available carrier bandwidth, number of aggregated carriers, transmit power, and the number of network nodes or base stations. The offered results highlight that only two carriers suffice to substantially facilitate resolving the integer ambiguity problem {while also largely enhancing the robustness of positioning against imperfections imposed by the network-side clocks and multi-path propagation. In addition, we also propose a %computationally efficient
 two-stage practical estimator that achieves the derived bounds 
under all realistic bandwidth and transmit power conditions. Furthermore, we show that with an additional search-based refinement step, the proposed estimator becomes particularly suitable for narrowband Internet of Things (IoT) applications operating efficiently even under narrow carrier bandwidths. Finally, both the derived bounds and the proposed estimators are extended to scenarios where the bands assigned to each base station are nonuniform or fully disjoint, enhancing the practical deployment flexibility.} 
%
%Carrier phase-based positioning (CPP) enables precise ranging, achieving sub-meter to centimeter accuracy, and has been widely used in global navigation satellite systems (GNSS) for localization and positioning purposes. Currently, it is widely used in cellular networks after combining phase and time measurements, performing differential phase measurements, and ensuring phase consistency across signals from different directions.  In this work, we extend the CPP model to a  multi-band scenario that supports intraband and interband carrier aggregation methods, considering multiple frequency bands such as FR1, FR2, and FR3 and we compare the fundamental performance bounds with benchmark mechanisms to demonstrate the superiority of  multi-band positioning systems.
\end{abstract}
\vspace{-1mm}

\vspace{-6mm}
\section{Introduction}
\label{section:sec1}
\textls[-2]{The 3\textsuperscript{rd} Generation Partnership Project (3GPP) Release 17 and 18 %(Rel-17/18) 
standards are setting stringent positioning requirements, targeting centimeter-level accuracy \cite{3gpp_tr_38_859_2024}, which makes centimeter-level positioning a key feature in next-generation terrestrial mobile communication systems. \emph{Carrier phase positioning (CPP)} is known to enable precise ranging, facilitating accuracies from sub-meter down to centimeter levels \cite{10475845}. CPP methods are already well-established in the global navigation satellite system (GNSS) context, %and widely used for localization and positioning 
%\cite{ding2021carrier}, 
{via either precise point positioning (PPP) or real-time kinematic 
(RTK) methods \cite{teunissen2015review}}, while the use of carrier phase measurements has recently been explored also in 3GPP standardization, particularly in the context of 5G-Advanced, as an emerging technology for enhanced positioning capabilities \cite{10536135}. {Unlike other range-based positioning methods and the related delay measurements, whose accuracy depends on the signal bandwidth \cite{li2004super, gezici2005localization}, carrier phase measurements are independent of the signal bandwidth, whether obtained using a phase locked loop (PLL) \cite{axelrad2002snr, mo2018novel} or classical estimators \cite{10437902}.}}

\vspace{-4mm}
\subsection{Fundamentals and Prior Art}
\textls[-4]{In general, the core principle of CPP lies in accurately measuring the carrier phase, 
%relative to a reference, 
enabling ranging precision within fractions of a wavelength \cite{10475845}.
A key technical challenge in utilizing carrier phase measurements is the \textit{integer ambiguity problem} \cite{10232971} which results from the fact that the carrier phase does not reveal the integer number of full wavelengths between the transmitter and the receiver. 
%In other words, while the phase can be measured with high precision, the total distance remains ambiguous due to the unknown integer multiple of the wavelength \cite{10232971}. 
Furthermore, in most systems, achieving strict time synchronization between the nodes is challenging, and clock biases can impede accurate integer ambiguity estimation in single-point positioning \cite{10522952}. The integer ambiguity problem can be addressed by combining phase and time measurements, or performing double-differential phase measurements across multiple base station (BS) pairs %, and ensuring phase consistency across signals from different directions 
\cite{8955972}. {On the other hand, synchronization can be directly addressed through inter-BS pilot transmission and clock bias estimation methods, which enable highly accurate synchronization on the order of picoseconds \cite{wu2024location, 9566601}.}
In the context of 5G New Radio (NR) mobile networks, the CPP has been recently studied in \cite{3gpp_tr_38_859_2024,9601204,9566601}, with main focus on single-band measurements and signal processing algorithms. % Unlike these existing studies, 
Additionally, \cite{10437902} introduced and derived performance bounds for localizing a user equipment (UE) under {a common} clock bias using \ac{TOA} and phase measurements, focusing again on % from multiple BSs over 
the single-band scenario.}

%Meanwhile, 
\textls[-12]{In parallel, carrier aggregation (CA) is a key technology in modern mobile networks, enabling the aggregation of fragmented spectrum resources \cite{10353006} for increased data rates. The NR CA mechanism applies to FR1 (sub-6 GHz) and FR2 (24.25–71 GHz), supporting various band combinations. %3GPP Release 16 later 
Additionally, new spectrum allocations are expected for 6G networks within the 7–24 GHz range (FR3), which balances wider bandwidth, lower attenuation, and reduced hardware costs \cite{6Gvision}. %\cite{cui20236g}. %It turns out that 
Importantly, CA can improve also the achievable positioning accuracy as it enlarges the effective bandwidth by combining positioning reference signals (PRS) from multiple component carriers %, effectively creating a signal with significantly higher overall bandwidth %—potentially spanning an entire frequency band 
\cite{9569341}. To this end, \cite{8746400} proposed joint estimation of  \ac{TOA} using reference symbols obtained from multiple frequency bands. Similarly, \cite{10285442} proposed a CA-based integrated sensing and communication mechanism %which  aggregates high and low-frequency bands 
for improved sensing performance. {Furthermore, in the context of GNSS, multi-band CPP is harnessed through dual- and triple-band measurements. For the global positioning system (GPS), these typically employ the civilian bands L1, L2, and L5, enabling more effective ionospheric error correction and improved positioning accuracy \cite{cocard2008systematic, feng2008gnss, li2012optimal, zhao2015three}. Finally, recent studies on multiple-frequency CPP in cellular networks \cite{li2022carrier, fan2023triple} exploit the carrier phases of closely spaced OFDM subcarriers within a single band, where combined phase measurements are used to facilitate integer ambiguity resolution. Since this approach is  effective for subcarriers that are very close in frequency, it cannot be readily extended to CA-based true multi-band scenarios. Moreover, existing CPP studies do not present sufficiently tight performance benchmarks or fundamental bounds reflecting the true potential of multi-band CPP,
%useful to evaluate the accuracy of the proposed estimators, 
while also lack practical estimation algorithms able to reach the realistic tight bounds.   }  }

% \textls[-2]{In this work, we extend the single-band CPP work in \cite{10437902}, and investigate the \emph{multi-band CPP} problem illustrated conceptually in Fig.~\ref{fig:system-model}. %In line with the support for multiple frequency-band combinations in
% We cover multi-band phase measurements in both intra-band and inter-band CA scenarios, while considering all the three frequency ranges (FR1, FR2, and FR3) that are of interest in 3GPP. We address the multi-band CPP from the fundamental performance bounds perspective, and the key technical contributions of this work can be stated  as follows:}

\vspace{-3mm}
\subsection{Novelty and Contributions}
\begin{figure}[t!]
		\centering
		\vspace{-1mm}
        \includegraphics[width=8.5cm]{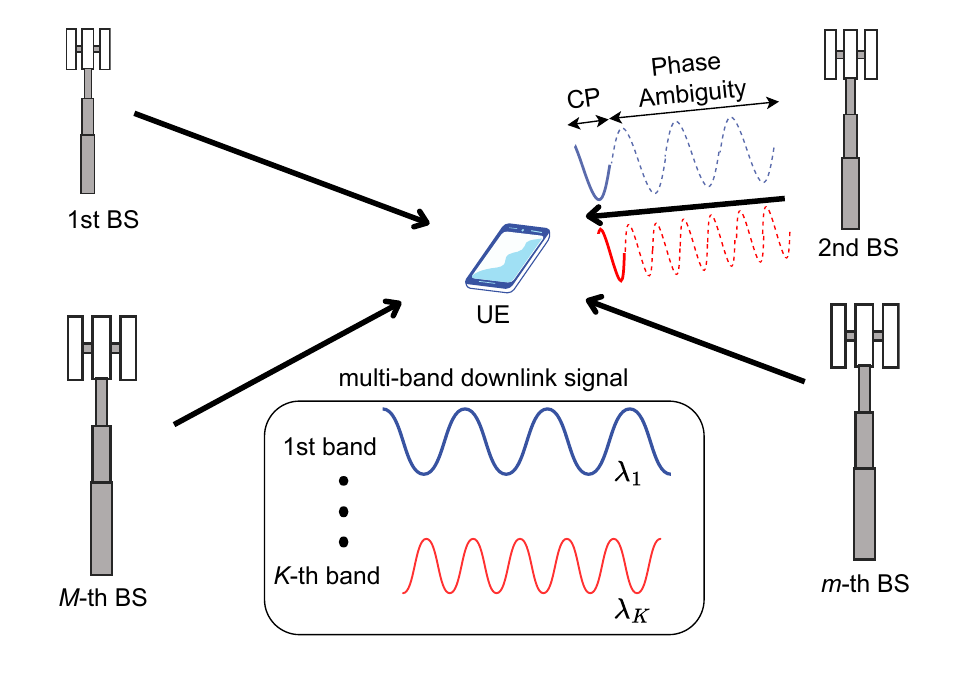}
        %SystemModelv3.pdf
		\vspace{-4mm}
        \caption{Illustration of the multi-band CPP system model with $K$ bands and $M$ network nodes.}
        \vspace{-3mm}
		\label{fig:system-model}
\end{figure}

\textls[-2]{{In this article, we investigate the {multi-band CPP} problem illustrated conceptually in Fig.~\ref{fig:system-model}. Specifically, while targeting to fill the above-noted gaps in the existing literature, 
%Building on the single-band analysis in \cite{10437902}, 
we derive the fundamental performance bounds for both intra- and inter-band CA scenarios across all three frequency ranges (FR1, FR2, and FR3) in terms of the mixed-integer Cramér-Rao bound (MICRB). We develop and formulate the related multi-band observation models, while also design and propose practical efficient estimators shown to be able to reach the derived bounds, and analyze their feasibility and performance under different system configurations. The key technical contributions and scientific novelty of this work can be stated as follows:}}

\begin{itemize}
%(i) We develop a signal model that characterizes the emergence of integer ambiguity when transitioning from waveform observations to delay and phase measurements in a multi-band downlink mechanism with multiple BSs. 
%(i) 
\item We formulate a model for the delay and phase observations in multi-band multiple-BS downlink transmission context, explicitly highlighting the integer ambiguity problem.
%(ii) 
%\item 
Based on this model, we introduce and derive the MICRB, specifically tailored for the considered multi-band observation model. This provides the fundamental position error bound (PEB) for cellular multi-band CPP; 
%These bounds offer valuable analytical insights, reducing the reliance on time-consuming simulations. 
%(iii) 
\item \textls[0]{{We propose a multi-band CPP %computationally tractable
estimation algorithm that is efficient in performance and achieves the derived bounds under all realistic configurations in terms of transmit power and carrier bandwidth. Unlike the methods in \cite{li2022carrier, fan2023triple}, the proposed algorithm is not restricted to specific frequencies or subcarriers and operates effectively for any aggregation of bands. In addition, we develop a search-based variant of the algorithm that preserves estimator efficiency even in low-bandwidth deployments and applications such as the Internet of Things (IoT);}}
%\item \textls[0]{\textcolor{blue}{We further develop a search-based variant of the algorithm suitable for low-bandwidth applications such as the Internet of Things (IoT);}}
\item We address the robustness of the proposed CPP methods against practical imperfections, covering mutual clock impairments between the network nodes as well as multipath propagation. We show that the multi-band operation allows for largely increased robustness against the considered practical impairments.
\item \textls[0]{{We investigate the feasibility of CPP and the proposed algorithms across different scenarios, where each BS operates on a potentially different subset of bands, while also highlight the importance of calibrated phase offsets among the bands through this analysis.}}
%\item \textls[0]{We offer a large collection of numerical results, covering the impacts of the carrier bandwidth, number of aggregated carriers, and the number of network nodes or BSs;} 
\end{itemize}
%Also, we compare the derived bounds with benchmark mechanisms and the single-carrier fusion bound to highlight the advantages of  multi-band CPP cellular positioning mechanism.
{{Overall, we utilize the derived bounds and the numerical results to gain and offer valuable insights into how multi-band operation affects the feasibility of  integer ambiguities resolution and the effective use of carrier-phase measurements for future cellular positioning.}

\textls[-4]{The rest of this article is organized as follows. Section \ref{section:sec2} presents the system model. In Section \ref{section:sec3}, we introduce the multi-band observation model tailored for the CPP problem and integer programming, derive the fundamental multi-band bounds, and extend them to scenarios with network-side synchronization impairments. Section \ref{section:sec4} proposes a two-stage CPP estimator along with its search-based variant, concluding with a computational complexity analysis. In Section \ref{section:sec5}, the proposed estimator is extended to the case of nonuniform BS-band assignment. Section \ref{section:sec6} presents and analyzes the numerical results, where the bounds and estimator performance are examined under various parameter settings, and the tightness of the bounds is verified. Sensitivity and robustness analyses against network-side clock impairments and multi-path propagation are also provided. Conclusions are drawn in Section \ref{section:sec7}, while selected mathematical details are given in the Appendices.} }

\textit{Notations:~}\textls[-10]{Italic letters $x$ denote scalars, while lowercase $\bm{x}$ and uppercase $\bm{X}$ in boldface represent vectors and matrices, respectively. The superscript $\left(\right)^\top$ denotes the transpose operation. The operator $\norm{\cdot}$ denotes the Euclidean norm, $|\cdot|$ denotes either the absolute value of a number or the cardinality of a set, and $\lfloor x \rfloor$ denotes the greatest integer less than or equal to $x$.   The operator $\otimes$ denotes the Kronecker product, $\diag{\bm{x}}$ outputs a diagonal matrix with the elements of a vector $\bm{x}$ on the diagonals, ${\rm blkdiag}({{\cdot,\cdot}})$ does the same thing for multiple matrix inputs and generates a block diagonal matrix, $\nabla_{\bm x}(\cdot)$ is the Jacobian matrix of its operand wrt. $\bm x$, and ${\rm{tr}}(\bm X)$ refers to the trace of $\bm X$.}

%Furthermore, FR3 empowers integrated sensing and communication (ISAC) applications by providing balanced frequency characteristics that avoid extreme Doppler effects and enable efficient multi-band sensing, leveraging existing radar technologies for enhanced angle, range, and detection capabilities.

%$\top$

%\section{Multi-band CPP Mechanism}
\vspace{-3mm}
\section{Multi-band CPP System Model}
\label{section:sec2}
We consider a scenario with a single UE and $M \geq N_d + 1$ BSs, where $N_d \in \{ 2,3\}$ represents the dimension of the position coordinates (i.e., 2D or 3D positions, respectively). The UE has an unknown position ${\bm{x}}_{\rm ue} \in \mathbb{R}^{N_d}$ and is subject to clock bias wrt.
%not perfectly time-synchronized to 
the BSs. We further assume that the BSs are mutually time- and phase-synchronized and have known locations ${\bm{x}}_{{{\rm bs}},m} \in \mathbb{R}^{N_d}$ \cite{10437902}. The impact of BS-BS synchronization imperfections is separately covered later in Section III.D. %Furthermore, the system supports coherent multi-band processing over $K$ different bands. 
%If time or phase synchronization does not hold, a reference station (RS) can be used to compute differential measurements, eliminating the per-BS time and phase biases.
Terminology-wise, we note that we use the notion of 'BS', however, the system model is also applicable with distributed transmission/reception points (TRPs) of a single BS -- or similarly with TRPs or antenna points of a future cell-free system.
Moreover, for simplicity, we assume that any frequency offsets among all entities have been corrected.

At the $k$-th band, each BS transmits a unit-modulus OFDM reference signal for $M$ OFDM symbols, e.g., 5G PRS, over $N_k$ subcarriers with subcarrier spacing $\Delta_{{ f},k}$ and transmit power  $P_{{\rm tx},k}$. 
%For convenience, we assume these parameters are independent of both the band and BS, and we omit the subscript $k$. 
The reference signal transmission of the $M$ BSs are orthogonalized, with each BS utilizing different subcarriers per OFDM symbol following the classical comb structure. %, with the transmitted power  $P_{{\rm tx},k}$.
%After the baseband and OFDM processing 
At the UE side, the frequency-domain received signal from $m$-th BS and $k$-th band under the ideal line-of-sight (LoS)\footnote{Impact of non-line-of-sight propagation paths is explicitly considered along the numerical results section, as part of the robustness analyses.} reads \cite{wymeersch2022radio}
\begin{equation}
 {\bm{y}}_{m,k} = \sqrt{E_{s,k}}  \alpha_{m,k}{\bm{d}}(\tau_{m,k}) +{{\bm{\omega}}_{m,k}},
\label{eq:eq2}
\end{equation}
such that ${\bm{y}}_{m,k}\in \mathbb{C}^{N_k\times 1}$. 
Herein, $E_{s,k}=P_{{\rm tx},k}/(N_k\Delta_{{ f},k})$ is the energy per subcarrier, ${{\alpha}}_{m,k}=\rho_{m,k}e^{{\color{black}-j2\pi\vartheta_{m,k}}}$ represents the complex channel gain in which $\rho_{m,k}$ captures the effect of path loss and transmitter and receiver antenna gains, {$\vartheta_{m,k}$ is the normalized carrier phase (expressed in cycles),} $\tau_{m,k}$ is the  \ac{TOA} {(delay)}, and 
$\bm d(\tau_{m,k})$ is the  channel response of the $m$-th BS for the $k$-th band with
$[{\bm{d}}(\tau_{m,k})]_n=e^{-j 2\pi n \Delta_{{ f},k} \tau_{m,k}}$
%$[{\bm{d}}(\tau_{m,k})]_n=e^{-j 2\pi n \Delta_{{ f},k} \tau_{m,k}}$, 
for $n\in\{0,\cdots,N_k-1\}-(N_k-1)/2$, and $k \in \{1,\dots,K\}$. The additive white Gaussian noise (AWGN) $\bm\omega_{m,k}\in\mathbb{C}^{N_k\times 1}$ has the statistics $\bm \omega_{m,k} \sim \mathcal{N}(\bm{0},\sigma_{\omega}^2\bm I)$, where $\sigma_{\omega}^2$ stands for the noise power spectral density\footnote{This is an assumption made for simplicity, but everything can be easily generalized for the case where $\sigma^2_\omega$ depends on $k$}. Moreover, the noise vectors for different bands or BSs are assumed to be mutually independent. 
%{\color{blue}Although the above formulation is expressed under the LoS assumption for notational simplicity, the bounds and estimators that will be derived in subsequent sections will also remain valid in multi-path environments, as the LoS path can be reliably identified. In particular, work \cite{8957110} has explicitly demonstrated this for the CPP scenario by employing the well-established ESPRIT algorithm. A multi-path sensitivity analysis will be presented in the numerical results.}
 
Since the BSs are mutually synchronized, the actual \ac{TOA} and carrier phase values across multiple frequency bands and BSs characterize the geometric relationship between the UE and the BSs. This can be expressed as \cite{esa2013gnss}
\begin{align}
    \tau_{m,k} & = \frac{1}{c}\| \bm x_{{{\rm{bs}}},m} - \bm x_{{\rm{ue}}}  \| + B_{\rm ue} \\
    %\vartheta_{m,k} & = \frac{2\pi}{\lambda_k}\| \bm x_{{{\rm{bs}}},m} - \bm x_{{\rm{ue}}}  \| + 2\pi f_{c,k} B_{\rm ue} + \phi_{{\rm ue},k} 
    \vartheta_{m,k} & = \frac{1}{\lambda_k}\| \bm x_{{{\rm{bs}}},m} - \bm x_{{\rm{ue}}}  \| +  f_{c,k} B_{\rm ue} + \varphi_{{\rm ue},k} 
\end{align}
%\textcolor{blue}{To Ehsan: Emphasize why Bue appears in Eq.4 and Eq.5 in this model} 
\textls[-1]{where $\tau_{m,k}$ is the actual \ac{TOA} in the time domain, and $\vartheta_{m,k}$ is the actual carrier phase in cycles. Here, $B_{\rm ue}$ is the clock bias of the UE {relative to the network/BSs clock} that is propagated to both the delay and carrier phase measurements,  $\varphi_{\rm ue,k}$ is the UE phase offset (in cycles), assumed to be band-dependent for generality, and $f_{c,k}=c/\lambda_k$ is the carrier frequency of the $k$-th band. Furthermore, $c$ denotes the speed-of-light and $\lambda_k$ is the wavelength corresponding to $f_{c,k}$.}

\textls[-1]{The considered multi-band CPP approach follows a two-stage estimation process. 
%similar to the single-band model presented in \cite{10437902}. 
First, the  \ac{TOA} and carrier phase are estimated from $\bm y_{m,k}$, expressed conceptually as $\hat{\tau}_{m,k}$ and $\hat{\vartheta}_{m,k}$. Then, these estimates are used to determine the UE position. Note that $\hat{\vartheta}_{m,k}$ is an estimate of the fractional part of $\vartheta_{m,k}$. Hence, we can express the  \ac{TOA} and phase measurements 
as effective distances $y_{\tau,m,k} = \hat{\tau}_{m,k}\times c$ and $y_{\vartheta,m,k} = \hat{\vartheta}_{m,k}{\lambda_k}$, given by}
\begin{align}
\label{toa measurements}
     y_{\tau,m,k} & = \| \bm x_{{\rm{bs}},m} - \bm x_{{\rm{ue}}}  \| + B_{{\rm ue}}c + \omega_{\tau,m,k} \\
    \begin{split}
    \label{phase measurements}
       y_{\vartheta,m,k} & = \| \bm x_{{\rm{bs}},m} - \bm x_{{\rm{ue}}}  \| +B_{\rm ue}c+  {z_{m,k}\lambda_{k}} \\&\quad +\varphi_{{\rm ue},k}{\lambda_k}+\omega_{\vartheta,m,k}.
    \end{split} 
\end{align}
Herein, $z_{m,k}\in \mathbb{Z}$ is the \emph{unknown integer ambiguity}, {with the actual value given by
\begin{equation}
    z_{m,k} = -\Big\lfloor \vartheta_{m,k}+ \frac{\omega_{\vartheta,{m,k}}}{\lambda_k} \Big\rfloor.
\end{equation}
The terms $\omega_{\tau,{m,k}}$ and $\omega_{\vartheta,{m,k}}$ represent zero-mean Gaussian noise affecting the \ac{TOA} and phase measurements or observations, respectively. Both terms are expressed in the distance domain, and are independent of each other. By deriving the Fisher information matrix (FIM)  for the physical signal model in (\ref{eq:eq2}) wrt. %with respect to 
$\tau_{m,k}$ and $\vartheta_{m,k}$, and applying a selected numerical approximation similar to \cite{10437902}, it can be shown that the lower bounds for the covariance  of $\omega_{\tau,{m,k}}$ and $\omega_{\vartheta,{m,k}}$ are equal to 
\begin{align}
    \label{eq:error_time}
    &{\sigma}_{\tau,m,k}^2 = { 3c^{ 2}/({2}\,{{\gamma}}_{m,k}\pi^{ 2} W_k^{\rm 2})} , \\
    \label{eq:error_phase}
    &{\sigma}_{\vartheta,m,k}^2 = {\lambda_{ k}^{2}/( 8\,{{\gamma}}_{ m,k}\pi^{ 2}) },
\end{align}
respectively. \textls[-2]{Herein, ${{\gamma}}_{m,k}$ is the signal-to-noise-energy ratio -- referred to as SNR in the continuation. This can be expressed as}
\begin{equation}
\label{eq:SNR}
    {{\gamma}}_{m,k} = N_k E_{s,k}\rho_{m,k}^2/\sigma_{\omega}^2
\end{equation}
% where $\sigma_{\omega}^2$ refers to XXX, %\forall k \,=\,1,\cdots,K \ ,  \ m = 1,\cdots,M$, 
where $W_k=N_k\Delta_{{ f},k}$ denotes the available passband width (available bandwidth) at band $k$.}

Finally, by gathering the observations from different BSs and frequency bands, we define $\bm{y}=\left[\bm{y}_{\tau}^\top , \bm{y}_{\vartheta}^\top\right]^\top$ where both ${\bm{y}}_{\tau}$ and ${\bm{y}}_{\vartheta}$ are structured as
${\bm{y}}_{\star}=\left[y_{\star,1,1}, y_{\star,2,1}, \dots, y_{\star,M,K}\right]^\top \in \mathbb{R}^{KM\times 1}$. Similarly, we build the effective noise vectors $\bm \omega_\tau$, $\bm \omega_\vartheta \in \mathbb{R}^{\rm{KM}\times \rm 1}$ by gathering $\omega_{\tau,{m,k}}$ and $\omega_{\vartheta,{m,k}}$, respectively. Their corresponding diagonal covariance matrices are defined as $\bm{\Sigma}_{\tau} = \diag{[ {\sigma}_{\tau,1,1}^2, {\sigma}_{\tau,2,1}^2,\ldots, \sigma_{\tau,M,K}^2 ]^\top} $, and $\bm{\Sigma}_{\vartheta} = \diag{[ {\sigma}_{\vartheta,1,1}^2, {\sigma}_{\vartheta,2,1}^2,\ldots, \sigma_{\vartheta,M,K}^2 ]^\top}$, respectively.
We further introduce  $\bm \eta = \left[\bm{s}^\top, \bm{z}^\top\right]^\top$ $\in \mathbb{R}^{(K+{N_d}+1)\times 1}\times \mathbb{Z}^{KM\times 1}$ as the \emph{unknown parameter vector}, where ${\bm{z}}=\left[z_{1,1}, z_{2,1}, \dots, z_{M,K}\right]^\top$ and ${\bm{s}}=\left[{{\bm{x}}}_{{\rm ue}}^\top, {B}_{\rm ue}, \bm{\varphi}_{{\rm ue}}^\top \right]^\top\in \mathbb{R}^{(K+{N_d}+1)\times 1}$, while $\bm{\varphi}_{{\rm ue}}=\left[\varphi_{{\rm ue},1}, \dots, \varphi_{{\rm ue},K}\right]^\top$. 

% \textls[-4]{By deriving the Fisher information matrix (FIM)  for the observation model (\ref{eq:eq2}) wrt. %with respect to 
% $\tau_{m,k}$ and $\vartheta_{m,k}$, and applying a selected numerical approximation \cite{10437902} {\color{blue}, it can be shown that the lower bounds for the covariance matrices of the measurement noises $\bm \omega_{\tau}$ and $ \bm \omega_{\vartheta}$, are diagonal matrices 
% $\bm \Sigma_{\vartheta}$ and $\bm \Sigma_{\tau}$,  respectively, where the diagonal elements corresponding to the $m$-th BS and $k$-th band are}
% \begin{align}
%     \label{eq:error_time}
%     &[\bm{\Sigma}_{\tau}]_{\ell(k,m),\ell(k,m)} = { (3c)^{ 2}/({2}\,{\rm{SNR}}_{m,k}\pi^{ 2} W_k^{\rm 2})} , \\
%     \label{eq:error_phase}
%     &[\bm{\Sigma}_{\vartheta}]_{\ell(k,m),\ell(k,m)} = {\lambda_{ k}^{2}/( 8\,{\rm{SNR}}_{ m,k}\pi^{ 2}) }.
% \end{align}}
%  where $\ell(k,m)={M \times (k-1) + m}$. Herein, ${\rm{SNR}}_{m,k}$ is the singal to noise energy ratio obtained by
% \begin{equation}
% \label{eq:SNR}
%     {\rm{SNR}}_{m,k} = N_k E_{s,k}\rho_{m,k}^2/\sigma_{\omega}^2
% \end{equation}
% % where $\sigma_{\omega}^2$ refers to XXX, %\forall k \,=\,1,\cdots,K \ ,  \ m = 1,\cdots,M$, 
% where $W_k=N_k\Delta_{{ f},k}$ denotes the available passband bandwidth at band $k$.}

%\section{Proposed Multi-Band Mixed-Integer Bound}
\vspace{-1mm}
\section{Multi-band CPP Performance Bounds} 
\label{section:sec3}
{\color{black} In this section, we introduce and derive a novel performance bound for the multi-band CPP problem that explicitly accounts for the mixed-integer nature of the observations -- thus referred to as the MICRB. Alongside the MICRB, we establish a multi-band delay-only bound, which is derived using more ordinary FIM formulations. %Although simpler, 
This bound provides useful insights and serves as a benchmark for comparison purposes. Furthermore, the bounds are also extended to account for practical synchronization impairments between the BSs, allowing for further insight and design requirements for future distributed networks.
%, which are modeled in the last subsection. 
%To initiate the derivation of the bounds, in the next subsection, we rewrite the observation equations in (\ref{toa measurements}) and (\ref{phase measurements}) in a more compact and suitable vector form.}

\vspace{-2mm}
\subsection{Multi-band Observation Model}
We first organize and express the stacked delay and phase observations in (\ref{toa measurements})-(\ref{phase measurements}) as
\begin{equation}
    \bm{y} = \tilde{\bm f}(\tilde{\bm s}) + \bm{B}\bm{z}+\bm{B}{\bm{\varphi}} + \bm{\omega},
    \label{eq:obs_model}
\end{equation}
% where ${\tilde{\bm{s}}}=\left[{{\bm{x}}}_{{\rm ue}}^\top, {B}_{\rm ue} \right]^\top$, and $\bm{\varphi} = {\bm \varphi}_{\rm ue}\otimes \bm{1}_{M\times 1} \in \mathbb{R}^{ KM\times 1}$. Additionally, The measurement noise is stacked as $\bm{\omega}=\left[\bm{\omega}_{\tau}^\top , \bm{\omega}_{\vartheta}^\top\right]^\top$   $\bm{\omega} \sim \mathcal{N}(\bm{0}_{2KM\times 1},\bm\Sigma_{\rm ch})$, $\bm\Sigma_{\rm ch} = {\rm blkdiag}(\bm\Sigma_{\tau},\bm\Sigma_{\vartheta})$, and $\bm{B} = \begin{bmatrix}\bm{0}_{KM\times KM} & \bm{\Lambda}\end{bmatrix}^\top \in {\mathbb{R}}^{ 2KM \times {KM}}$,  where  $\bm{\Lambda} = {\rm{diag}}([\lambda_1,\cdots,\lambda_K]^\top)\otimes{\bm I}_{{M}}\in {\mathbb{R}}^{ KM \times {KM}}$,
where ${\tilde{\bm{s}}}=\left[{{\bm{x}}}_{{\rm ue}}^\top, {B}_{\rm ue} \right]^\top$,  $\bm{\varphi} = \bm{\varphi}_{{\rm ue}}\otimes \bm{1}_{  M\times 1} \in \mathbb{R}^{   KM\times 1}$. The matrix $\bm{B}$ is defined as $\bm{B} = \begin{bmatrix}\bm{0}_{  KM\times KM} \!&\! \bm{\Lambda}\end{bmatrix}^\top $  with  $\bm{\Lambda} = \text{diag}([\lambda_1,\cdots,\lambda_K]^\top)\otimes{\bm I}_{{M}}\in {\mathbb{R}}^{   KM \times {KM}}$. The noise is stacked as $\bm{\omega}=\left[\bm{\omega}_{\tau}^\top , \bm{\omega}_{\vartheta}^\top\right]^\top$ with   $\bm{\omega} \sim \mathcal{N}(\bm{0}_{  2KM\times 1},\bm\Sigma_{\rm ch})$, and $\bm\Sigma_{\rm ch} = {\rm blkdiag}(\bm\Sigma_{\tau},\bm\Sigma_{\vartheta})$,
while $\tilde{\bm f}(\cdot)$ is a nonlinear function of $\tilde{\bm s}$, defined for $j = 1,2,\cdots, 2KM$ as
\begin{align}
    [\tilde{\bm f}(\tilde{\bm s})]_j &= \|\bm x_{{\rm bs},{\rm mod}(j-1,M)+1 } - \bm x_{{{\rm ue}}} \| + B_{\rm ue}c.
    %\\ [\tilde{\bm f}(\tilde{\bm s})]_{KM+j} &= \|\bm x_{{\rm bs},{\rm mod}(j,M) } - \bm x_{{\rm ue}} \| 
\end{align}

We next note that $\bm \varphi$ and $\bm z$ are not jointly identifiable, and the system of equations is under-determined. 
To address this, since both $\bm z$ and $\bm \varphi_{\rm ue}$ are eventually nuisance parameters, %for each band, 
the state dimension can be reduced by incorporating $\varphi_{{\rm ue},k}$ into one of the integer ambiguities of the corresponding band. Without loss of generality, if we merge each $\varphi_{{\rm ue},k}$ with $z_{1,k}$, the reduced-state variables for the $k$-th band become $\varphi_{{\rm d},k} = {\varphi_{{\rm ue},k}} + z_{1,k}$ and $[z_{2,k} - z_{1,k}, \dots, z_{M,k} - z_{1,k}]^\top \in \mathbb{Z}^{(M-1) \times 1}$, instead of the original $\varphi_{{\rm ue},k}$ and $[z_{1,k}, \dots, z_{M,k}]^\top \in \mathbb{Z}^{M \times 1}$. Accordingly, the observation model can be reformulated using the combined parameters as
\begin{equation}\label{eq:diff}
    \bm{y} = \tilde{\bm f}(\tilde{\bm s}) + \bm{B}\bm{E}\bm{z}_{\rm d}+\bm{B}\bm{\varphi}_{\rm d} + \bm{\omega},
\end{equation}
\textls[-2]{where $\bm z_{\rm d} = \bm D \bm z \in \mathbb{Z}^{K(M-1)\times 1} $ denotes the multi-band differential (reduced-dimension) integer ambiguity vector, and the combined phase offset vector is $\bm{\varphi}_{\rm d} = \bm{\varphi}_{\rm d,ue}\otimes \bm{1}_{M\times 1} \in \mathbb{R}^{ KM\times 1}$ with $\bm{\varphi}_{\rm d,ue} = [\varphi_{{\rm d},1},\cdots,\varphi_{{\rm d},K}]^\top$. Both $\bm{D}$ and $\bm{E}$ are block-diagonal matrices, with each block corresponding to a band. Matrix $\bm D =  \bm{I}_{K}\otimes\begin{bmatrix}-\bm{1}_{(M-1)\times 1} & \bm{I}_{(M-1)}\end{bmatrix} \in \mathbb{R}^{K(M-1)\times KM}$ acts as a differentiator, reducing the integer parameters dimension in each band. Furthermore, matrix $\bm E = \bm{I}_{K}\otimes\begin{bmatrix}
    \bm{0}_{ (M-1)\times 1 } & \bm{I}_{(M-1)}
\end{bmatrix}^\top\in \mathbb{R}^{KM \times K(M-1)}$ inserts a zero element as the first row of each reduced-dimension integer block.} 

Now, we can collect all the real-valued unknowns into ${\bm f}({\bm s}) = \tilde{\bm f}(\tilde{\bm s}) +\bm{B}\bm{\varphi}_{\rm d}$ and rewrite (\ref{eq:diff}) as 
\begin{equation}\label{eq:diff_}
    \bm{y} = {\bm f}({\bm s}) + \bm{B}\bm{E}\bm{z}_{\rm d} + \bm{\omega}.
\end{equation}
%where ${\bm f}({\bm s}) = \tilde{\bm f}(\tilde{\bm s}) +\bm{B}\bm{\varphi}_{\rm d}$.
This observation model serves as the basis of the MICRB derivation, \textcolor{black}{in which with a slight change of variables, $\bm{s}$ is redefined as ${\bm{s}}=\left[{{\bm{x}}}_{{\rm ue}}^\top, {B}_{\rm ue}, \bm{\varphi}_{\rm d,ue}^\top \right]^\top\in \mathbb{R}^{(K+{N_d}+1)\times 1}$. }

\vspace{-4mm}
\subsection{Multi-band Known-Integer and Mixed-Integer CRBs}
\vspace{-0mm}
{\subsubsection{Methodology} For conventional estimation problems with ordinary real- or complex-valued parameters, such as \ac{TOA}-based positioning, the lower bound on the error covariance of any unbiased estimator is obtained from the diagonal elements of the inverse of the FIM \cite{kay1993fundamentals}.
%Fisher Information Matrix (FIM). 
This is known as the Cramér-Rao Bound (CRB), 
%defining the minimum achievable variance for any unbiased estimator, 
while an {\color{black}unbiased estimator} whose covariance matrix equals the CRB is called an efficient estimator.
% I will use this definition "efficient estimator" later on.
%
In general, the FIM quantifies how sharply the likelihood function varies around the true parameter values—essentially, how much curvature the likelihood function exhibits. If the log-likelihood function of the observation vector $\bm o$ wrt. the parameter vector $\bm \theta$ is denoted as $L(\bm o ; \bm \theta)$, the FIM is given by ${\rm FIM} = \mathbb{E}[\nabla_{\bm \theta}L(\bm o ; \bm \theta)\, (\nabla_{\bm \theta}L(\bm o ; \bm \theta))^\top]\big|_{\bm \theta_{\rm true}}$ \cite{kay1993fundamentals}. 

However, in the CPP problem at hand, the conventional FIM approach cannot be applied since the above derivatives cannot be computed with respect to \emph{integer-valued} parameters. 
%Instead, the proposed methodology directly computes the MICRB without deriving the corresponding FIM. The approach can be summarized as follows: 
Instead, we derive and compute the so-called multi-band mixed-integer CRB as follows: 
We first assume that the relaxed integer ambiguities, considered as ordinary real numbers, are {\color{black}given} with the error covariance equal to the  CRB of the relaxed observation model. %(as definded in the M.kay estimation theory book, efficient estimator reaches the corresponding CRLB).  
Next, to incorporate the integer nature of the parameters into the bound derivation, we find the integer least squares (ILS) estimation of the ambiguities.  This transforms the remaining part of the problem into the \emph{real-valued} domain, referred to as the \emph{known-integer} problem.  However, any error in the estimated integer ambiguities from the ILS stage introduces a bias in  $\bm s$. Thus, in the final stage, we compute the MICRB as the lower bound for the error covariance of a biased estimator of $\bm s$, based on the theoretical framework presented in \cite{1658250}.} 

% \textls[-1]{As stated, we assume %efficient estimation of 
% the relaxed ambiguity variables %$\bm z$ 
% lie in the real-valued domain, with an error covariance equal to the CRB. %It is worth noting that if we intend to
% %To this end, when estimating $\bm z$ as a real-valued parameter, the nuisance parameter $ \varphi_{{\rm ue},k}$ will also be absorbed into $ z_{m,k}$.
% %To this end, when estimating $\bm z$ as a real-valued parameter, the nuisance parameter $ \varphi_{{\rm ue},k}$ will also be absorbed into $ z_{m,k}$.
% {\color{blue}Since  $\bm \varphi$ and $\bm z$ are not jointly identifiable, the relaxed ambiguities absorb $\bm \varphi$. Using the observation model \eqref{eq:diff} as the baseline, the relaxed model takes the form
% \begin{equation}\label{eq:diff_rlx}
%     \bm{y} = \tilde{\bm f}(\tilde{\bm s}) + \bm{B}\bm{z}_{\rm rlx}+\bm{\omega},
% \end{equation}
% where the relaxed integer ambiguity vector, ${\bm{z}}_{\rm rlx}\in \mathbb{R}^{KM\times 1} $ is modeled by} 
% %the estimation model  ${\bm{z}}_{\rm rlx}\in \mathbb{R}^{KM\times 1} $  is: 
% \begin{equation}\label{eq:relaxed_intg}
%     {\bm{z}}_{\rm rlx} = \bm{E}\bm{z}_{\rm d}+ \bm{\varphi}_{\rm d} + \bm{u} , \ \ 
%     \bm{u} \sim \mathcal{N}(\bm{0}_{KM\times 1},\bm\Sigma_{\rm rlx})
% \end{equation}
% with $\bm\Sigma_{\rm{rlx}}(\bm z)$ denoting the lower bound on the error covariance matrix of the relaxed integer unknowns. The detailed expression of this matrix is provided in Appendix \ref{apdx a}. 

\vspace{-3mm}
\textls[-14]{{\color{black}\subsubsection{Established Bounds} As stated, we start by assuming that the real-valued relaxed ambiguity variables are given, with an error covariance equal to the corresponding CRB. 
 Before proceeding, we ensure that all parameters are identifiable in the relaxed observation model.  Using  \eqref{eq:diff} as the baseline, the observation model with the relaxed identifiable parameters takes the form
\begin{equation}\label{eq:diff_rlx}
    \bm{y} = \tilde{\bm f}(\tilde{\bm s}) + \bm{B}\bm{z}_{\rm rlx}+\bm{\omega}.
\end{equation}
Through this, by computing the CRB of $\bm{z}_{\rm rlx}\in \mathbb{R}^{KM\times 1}$, denoted by $\bm{\Sigma}_{\rm rlx}$,  we can express the given real-valued ambiguities as follows
\begin{equation}\label{eq:relaxed_intg}
    \hat{\bm{z}}_{\rm rlx} = \bm{E}\bm{z}_{\rm d}+ \bm{\varphi}_{\rm d} + \bm{u} , \ \ 
    \bm{u} \sim \mathcal{N}(\bm{0}_{KM\times 1},\bm\Sigma_{\rm rlx}).
\end{equation}
The detailed expression of $\bm\Sigma_{\rm rlx}$ is provided in Appendix \ref{apdx a}.}}
Next, to eliminate the term $\bm{\varphi}_{\rm d}$ in $\hat{\bm{z}}_{\rm rlx}$, we multiply both sides of \eqref{eq:relaxed_intg} by $\bm D$ and obtain the differential observation 
\begin{equation}
    \bm r = \bm{D}\bm{E}\bm{z}_{\rm d}+ \bm{D}\bm{\varphi}_{\rm d} + \bm{D}\bm{u} = \bm{z}_{\rm d} + \bm{D}\bm{u},
\end{equation}
which follow from the equalities of $\bm{D}\bm{E} = \bm I$ and $\bm{D}\bm{\varphi}_{\rm d} = \bm 0$. {\color{black}To prove the first equality, note that both $\bm{D}$ and $\bm{E}$ are block-diagonal matrices, so their product is also block-diagonal, with each block satisfying $\begin{bmatrix}-\bm{1}_{(M-1)\times 1} & \bm{I}_{(M-1)}\end{bmatrix}\times\begin{bmatrix}
    \bm{0}_{ (M-1)\times 1 } & \bm{I}_{(M-1)}
\end{bmatrix}^\top = \bm{I}_{(M-1)}$. For the second equality, we know $\bm{\varphi}_{\rm d}$ consists of $K$ subvectors, the $k$-th one being $\varphi_{{\rm d},k}\times\bm{1}_{M\times 1} $. The block multiplication with $\bm{D}$ cancels the constant entries in each subvector, resulting in the zero vector. Finally, we recover $\bm{z}_{\rm d}$ by solving the ILS problem, written as 
\begin{equation}\label{integer-prog}
    \hat{\bm{z}}_{\rm d} =  \underset{\bm{z}_{\rm d}\in \mathbb{Z}^{K(M-1)\times 1}}{\arg \min}(\bm r  - \bm{z}_{\rm d})^\top \bm  S^{-1}  (\bm r  - \bm{z}_{\rm d}),
\end{equation}
where $\bm S = \bm D \bm\Sigma_{\rm rlx}\bm D$.}

In general, any %integer ambiguity 
error in $\bm{z}_{\rm d}$ {\color{black} leads to a bias in $\bm{s}$.} To determine this bias, we linearize (\ref{eq:diff_}) around $\bm s$, which yields $\bm{y} = \bm f (\bm s) +  \bm A_{\bm f}(\bm s)\bm \delta \bm s + \bm{B}\bm{E}\bm{z}_{\rm d} + \bm{\omega}$, where $\bm A_{\bm f}(\bm s) = \nabla_{\bm s}\bm f (\bm s) \in \mathbb{R}^{2KM\times {(K+{N_d}+1)}}$ is computed at the true value of $\bm s$. {\color{black}The explicit form of $\bm A_{\bm f}(\bm s)$ is provided in the next section, in Eq. \eqref{eq: gradient of s0}.
In the absence of any error in $\bm{z}_{\rm d}$, i.e., when $\hat{\bm{z}}_{\rm d} = \bm{z}_{\rm d}$, the WLS estimate of $\bm{\delta s}$ is given by
\begin{equation} \label{eq:deltas_est}
\widehat{\bm{\delta s}} = \big(\bm{\Sigma}_{\rm ch}^{-1/2} \bm A_{\bm f}(\bm s)\big)^\dagger \bm{\Sigma}_{\rm ch}^{-1/2} \left( \bm{y} - \bm{f}(\bm{s}) - \bm{B} \bm{E} \hat{\bm{z}}_{\rm d} \right),
\end{equation}
which is an unbiased\footnote{Strictly speaking, the unbiasedness holds under the linearized model. Because $\bm f$ is nonlinear, the estimator is approximately unbiased, but the approximation is highly accurate except when noise dominates. } zero-mean estimator.
However, if there is an integer error $\bm{\delta}_{z}$ in the solution to the integer problem \eqref{integer-prog}, such that $\hat{\bm{z}}_{\rm d} = \bm{z}_{\rm d} + \bm{\delta}_{z}$, the resulting estimate $\widehat{\bm{\delta s}}$ will have a bias given by
\begin{equation}\label{eq:propagated bias}
    \bm b (\bm s | \bm{\delta}_{z} ) = -\big(\bm \Sigma_{\rm ch}^{-1/2}\bm A_{\bm f}(\bm s) \big) ^\dagger \bm \Sigma_{\rm ch}^{-1/2} \bm B \bm E  \bm{\delta}_{z}
\end{equation}}
 Now, the observation model is turned into a known integer one,  %we can develop a (biased) estimator of $\bm{s}$. 
 meaning that there is no longer any integer unknown in the problem. In order to compute the MICRB, we first note that when the estimators of $\bm{s}$ are biased with  bias $\bm b (\bm s | \bm{\delta}_{z} )$, the corresponding {\color{black}lower bound of the error covariance}, accounting for such bias, is given by \cite{1658250}
\begin{equation}
\begin{split}
    \bm \Sigma_{\rm{mi}}(\bm s | \bm{\delta}_{z} ) = & \bm b (\bm s | \bm{\delta}_{z} )\bm b (\bm s | \bm{\delta}_{z} )^\top + \\ & \Big(\bm I + \bm A_{\bm b}(\bm s)\Big)\bm{\Sigma}_{\text{known}}(\bm{s})\Big(\bm I + \bm A_{\bm b}(\bm s)\Big)^\top
\end{split}
\end{equation}
%\begin{equation}
  %  \bm \Sigma(\bm s | \bm \delta ) = \bm b (\bm s | \bm \delta )\bm b (\bm s | \bm \delta )^\top + \bm{\Sigma}_{\text{known}}(\bm{s})
%\end{equation}
where $\bm A_{\bm b}(\bm  s) = \nabla_{\bm s}\bm b (\bm s | \bm{\delta}_{z} ) \in \mathbb{R}^{{(K+{N_d}+1)\times (K+{N_d}+1)}} $. In the usual case that the bias is not very sensitive to small variations of $\bm s$, we can approximate that $\bm A_b (\bm  s)\approx \bm 0$, and $\bm{\Sigma}_{\text{known}}(\bm{s})$ represents the \emph{ultimate lower bound} on the error covariance matrix for the measurement model in which the  {\color{black}reduced-state integer ambiguity vector, $\bm z_{\rm d}$},  is assumed to be known. The detail formulation of the matrix $\bm{J}_{\text{known}}(\bm s) = \bm{\Sigma}_{\text{known}}^{-1}(\bm s)$ can be found in Appendix \ref{apdx b}. The corresponding bound is called \emph{known-integer bound} in the following.

Finally, the proposed MICRB on the error covariance can be obtained by taking the expectation with respect to the integer ambiguity error. This is expressed as
\begin{equation}
\label{j_mixed}
    \bm{\Sigma}_{\text{mi}}(\bm s) = \mathbb{E}_{\bm{\delta}_{z}} \bigg[\bm \Sigma_{\rm{mi}}(\bm s | \bm{\delta}_{z} )\bigg]\approx \frac{1}{N_{\rm mc}}\sum_{i=1}^{N_{\rm mc}}\bm\Sigma_{\rm{mi}}(\bm s | \bm \delta_z^{(i)} ),
    %\sum_{\bm{\delta}_{z} \in \mathbb{Z}^{K(M-1)\times 1}} {\rm {Pr}}(\bm{\delta}_{z}) \bm \Sigma_{\rm{mi}}(\bm s |  \bm{\delta}_{z} ).
\end{equation}
\textls[-10]{where $N_{\rm mc}$ is the number of Monte-Carlo simulations, serving as the basis for numerical computations. 
%As shown in \eqref{j_mixed}, the expectation can be numerically computed using Monte-Carlo simulations.  
Since (\ref{integer-prog}) is invariant wrt. ${\bm{z}}_{\rm d}$, the Monte-Carlo approach simplifies to generating $N_{\rm mc}$ samples of a Gaussian noise, $\bm{r}^{(i)} \sim \mathcal{N}(\bm{0}_{K(M-1)\times 1},\bm S)$, and directly determining the integer error  $\bm\delta_z^{(i)} =\hat{\bm z}_{\rm d}^{(i)}$ using (\ref{integer-prog}). Hence, the floating solution of the original problem is actually not needed to obtain the MICRB. 
%This involves generating $N_{\rm mc}$ samples as $\bm r^{(i)} = \bm S^{-1/2}\bm{z}_{\rm d}+ \bm{u}^{(i)}$, where $\bm{u}^{(i)} \sim \mathcal{N}(\bm{0}_{K(M-1)\times 1},\bm I)$, and determining the estimate $\hat{\bm z}_{\rm d}^{(i)}$ using (\ref{integer-prog}). From this, we obtain $ \bm\delta^{(i)} = \hat{\bm{z}}_{\rm d}^{(i)} - {\bm{z}}_{\rm d}$. 
Finally, 
%$\bm{\Sigma}_{\text{mi}}(\bm s)\approx \frac{1}{N_{\rm mc}}\sum_{i=1}^{N_{\rm mc}}\bm\Sigma_{\rm{mi}}(\bm s | \bm \delta_z^{(i)} )$, and
$\bm{\Sigma}_{\text{mi}}(\bm x_{\rm ue}) = [\bm{\Sigma}_{\text{mi}}(\bm s)]_{1:{N_d},1:{N_d}}$. Thus, computing the MICRB requires only the geometry matrix $\bm{A}_{\bm f}(\bm s)$, along with $\bm{\Sigma}_{\rm{rlx}}$, $\bm{\Sigma}_{\rm{ch}}$, $\bm{\Sigma}_{\rm{known}}$. An overall holistic block diagram illustrating the proposed MICRB is depicted in Fig. ~\ref{fig:visual_abstract}.}
% \begin{figure*}[t!]
% 		\centering
% 		\includegraphics[width=9cm]{images/vis_abs1.png}
%         %SystemModelv3.pdf
% 		\vspace{-1mm}
%         \caption{\textcolor{blue}{Block diagram of the Mixed-Integer Bound derivation}}
%         \vspace{2mm}
% 		\label{fig:visual_abstract}
% \end{figure*}

\begin{figure*}[!t]
    \centering
    \begin{tikzpicture}[
        block/.style={rectangle, minimum height=0.8cm, minimum width=2.5cm, rounded corners, thick, font=\footnotesize},
        node distance=0.9cm and 0.9cm
    ]
        % Define color variables
        \definecolor{myorange}{rgb}{1.0, 0.8, 0.4}
        \definecolor{myorangedark}{rgb}{0.9, 0.4, 0.0}
        \definecolor{myblue}{rgb}{0.6, 0.8, 1.0}
        \definecolor{mybluedark}{rgb}{0.2, 0.4, 0.8}
        \definecolor{mygreen}{rgb}{0.67, 0.80, 0.60}
        \definecolor{mygreendark}{rgb}{0.0, 0.6, 0.0}
        \definecolor{deepred}{rgb}{0.6, 0.0, 0.0}

        \definecolor{lightgrayblock}{RGB}{220,220,220}   % very light gray
        \definecolor{darkgrayblock}{RGB}{120,120,120}    % darker gray

        % Upper layer (orange)
        \node[block, fill=white, draw=black] (u1) {Construct the observation model as in \eqref{eq:obs_model}};
        \node[block, fill=white, draw=black, right=of u1] (u2) {\shortstack{Reduce the dimension of nuisance parameters %to obtain the determined 
        to \\ obtain the model \eqref{eq:diff_} with identifiable unknowns}};

        % Lower layer (blue)
        \node[block, fill=white, draw=black, below=of u1] (l1) {\shortstack{
        %Generate the relaxed real-valued \\ ambiguities  from the model in
        Generate a random sample $\bm r \sim \mathcal{N}(\bm{0},\bm S)$}};
        \node[block, fill=white, draw=black, right=of l1] (l2) {\shortstack{Recover   $\bm \delta_z = \hat{\bm z}_{\rm d}$ from \eqref{integer-prog},\\ converting the measurements  \\ into a known-integer model}};
        \node[block, fill=white, draw=black, right=of l2] (l3) {\shortstack{Formulate the propagated bias\\to  $\bm  s$,   as described in \eqref{eq:propagated bias} }};
        
        % Final output (green)
        \node[block, draw=black, below=0.6cm of l2, ] (out) {{\color{black}\textbf{Multi-band Mixed-Integer Cramer-Rao Bound (MICRB)} \eqref{j_mixed}}};
        %$\bm{\Sigma}_{\text{mi}}(\bm s) = \mathbb{E}_{\bm{\delta}_{z}} \bigg[\bm \Sigma_{\rm{mi}}(\bm s | \bm{\delta}_{z} )\bigg]\approx \frac{1}{N_{\rm mc}}\sum_{i=1}^{N_{\rm mc}}\bm\Sigma_{\rm{mi}}(\bm s | \bm \delta_z^{(i)} )$  }};

        % Overlayed white equation block (adjust yshift for vertical positioning)
        % \node[draw=black, fill=white, rounded corners,  anchor=north west, minimum height=0.6cm, font=\footnotesize] at ([yshift=-0.1cm,xshift=7.4cm]out.north west)
        % {
        % $\bm{\Sigma}_{\text{mi}}(\bm s) = \mathbb{E}_{\bm{\delta}_{z}} \bigg[\bm \Sigma_{\rm{mi}}(\bm s | \bm{\delta}_{z} )\bigg]\approx \frac{1}{N_{\rm mc}}\sum_{i=1}^{N_{\rm mc}}\bm\Sigma_{\rm{mi}}(\bm s | \bm \delta_z^{(i)} )$
        % };

        % Orange arrows for upper layer
        \draw[thick, draw=black, -{Latex[length=2mm, width=2mm]}] (u1) -- (u2);

        % Orange to blue connection (orange arrow)
        \draw[thick, draw=black, -{Latex[length=2mm, width=2mm]}] (u2.east) -| (l3.north);

        % Blue arrows for lower layer
        \draw[thick, draw=black, -{Latex[length=2mm, width=2mm]}] (l1) -- (l2);

        % Blue arrows for lower layer
        \draw[thick, draw=black, -{Latex[length=2mm, width=2mm]}] (l2) -- (l3);

        % Green arrow for final output
        \draw[thick, draw=black, -{Latex[length=2mm, width=2mm]}] (l3.south) -- ++(0, -0.4) -| (out.north);

        % % Oval loop above Lower 1
        % \path let \p1 = (l1.north) in
        %     coordinate (top) at (\x1, \y1);
        % \draw[
        %     thick, draw=black,
        %     postaction={decorate},
        %     % decoration={
        %     %     text along path,
        %     %     text={|\scriptsize\sffamily|Do this for enough number of Monte-Carlo trials},
        %     %     text align=center,
        %     %     raise=2pt
        %     % },
        %     -{Latex[length=2mm, width=2mm]}
        % ] 
        % ([yshift=0cm]top) arc[start angle=-90, end angle=-450, x radius=1.8cm, y radius=0.7cm];

        % \node[
        %     align=center,
        %     font=\footnotesize\sffamily,
        %     text=black
        % ] 
        % at ([yshift=0.65cm]top) {Do this for enough number\\ of Monte-Carlo trials};
        \draw[thick, draw=black, -{Latex[length=2mm, width=2mm]}]
            (l1.east) 
            arc[start angle=-30, end angle=210, 
            x radius=3cm, y radius=0.5cm]
        -- (l1.west)
       node[pos=0.25, above, xshift=2.6cm, yshift=0.65cm, align=center] {\footnotesize
         Repeat for $N_{\rm mc}$ Monte-Carlo trials};]

    \end{tikzpicture}
    \vspace{-1mm}
    \caption{\textcolor{black}{Block diagram illustration of deriving and computing the multi-band mixed-integer Cramér-Rao bound.}}
    \vspace{-2mm}
    \label{fig:visual_abstract}
\end{figure*}
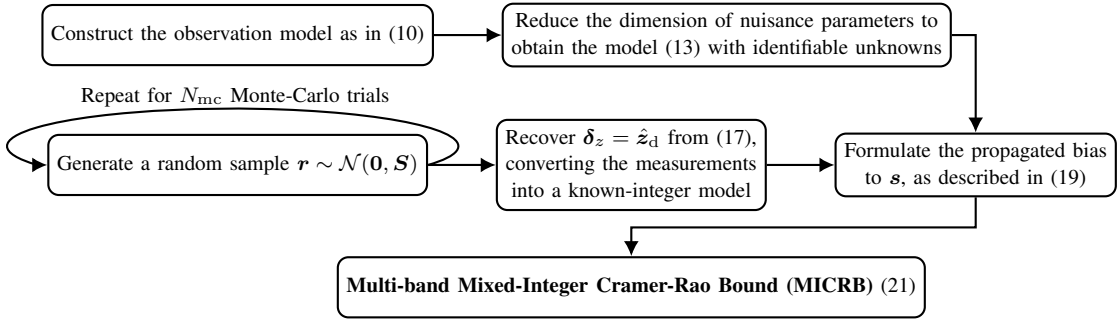

\vspace{-3mm}
\subsection{Multi-band Delay-Only Reference Bound}
As an important reference or benchmark, besides the known-integer bound, we consider the bound for the observation model that exploits only the  \ac{TOA} measurements and thus drops $\bm y_{\vartheta}$. 
In this case, the unknown parameters will be ${\tilde{\bm{s}}}=\left[{{\bm{x}}}_{{\rm ue}}^\top, {B}_{\rm ue} \right]^\top$, and the corresponding  FIM can be formulated as
\begin{equation}
\label{j_delay}
{\bm{J}}_{\text{delay}}(\bm{\tilde s}) = 
\begin{bmatrix}
   {\bm{\tilde{U}}} \bm{J}_{\tau} {\bm{\tilde{U}}}^\top& c \,{\bm{\tilde{U}}} \, \text{diag}(\bm{{J}}_{\tau})\\
   c ({\bm{\tilde{U}}} \, \text{diag}(\bm{{J}}_{\tau}))^\top & {\rm tr}({\bm{J}}_{\tau}) c^2 
\end{bmatrix},
\end{equation}
where  $\bm{J}_{\tau}=\bm\Sigma_{\tau} ^{ -1}$, and ${\bm{\tilde{U}}}={\bm 1}_{ {1\times K}}\otimes\bm{U}\in {\mathbb{R}}^{ N_d \times {KM}}$ 
{\color{black}denotes the transposed Jacobian matrix of $[\tilde{\bm f}(\tilde{\bm s} )]_{ 1:{KM}}$ wrt. $\bm x_{\rm ue}$,} 
with ${\bm{U}}=\left[  {\bm{u}}_1, {\bm{u}}_2,\dots, {\bm{u}}_M \right]$,  and $ {\bm{u}_{\rm m}} = ({\bm{ x}_{{\rm ue}}}-{\bm {x}_{{\rm bs},m}})/ \|{\bm x_{{\rm ue}}}-{\bm x_{{\rm bs},m}}\| $.\footnote{All $\bm{\tilde{U}}$, $\bm{U}$, and $\bm{u}_m$  depend on $\bm{x}_{\rm ue}$, but the argument $(\bm{x}_{\rm ue})$ is omitted for notational simplicity.} Then, we obtain $\bm{\Sigma}_{\text{delay}}({\bm{x}}_{\text{ue}})= \left[{\bm{J}}_{\text{delay}}^{\rm -1}(\bm{\tilde s})\right]_{ 1:{N_d},1:{N_d}} $.
 This lower bound is %\emph{pessimistic} and 
 greater in value than the known-integer bound since it completely disregards the phase measurements. Thus, the difference between the delay-only bound and the derived known-integer and mixed-integer bounds tells the potential performance gains available through the phase measurements.

{\color{black}
\vspace{-3mm}
\subsection{Extension to System with Network/BS Clock Imperfections}
\label{clock-imp}
While the above derivations assume perfect mutual synchronization among the BSs, achieving such in practice is challenging, and each {BS} may exhibit an individual residual clock bias relative to a reference BSs. While most existing works \cite{fouda2022toward, ou2024single} address this issue by employing an additional reference receiver and applying double-differential measurements, in our scenario these imperfections cannot be estimated from the available measurements nor eliminated through differential methods. Thus, to characterize their impact, we instead model the imperfections as random variables with given statistics.

To this end, we next extend the previous bounds to an imperfectly synchronized network by modeling BS-dependent clock imperfections as additional random terms, $\tilde{B}_m \sim \mathcal{N}(0,\delta_m^2)$ \cite{9566601}, added to both the \ac{TOA} and carrier-phase observation models in \eqref{toa measurements} and \eqref{phase measurements}. The modified observations become
\begin{align}
\label{toa measurements-imperfect}
     y_{\tau,m,k} & = \| \bm x_{{\rm{bs}},m} - \bm x_{{\rm{ue}}}  \| + B_{{\rm ue}}c + \tilde B_{m}c + \omega_{\tau,m,k} \\
    \begin{split}
    \label{phase measurements-imperfect}
       y_{\vartheta,m,k} & = \| \bm x_{{\rm{bs}},m} - \bm x_{{\rm{ue}}}  \| + {z_{m,k}\lambda_{k}} +\varphi_{{\rm ue},k}{\lambda_k}+ \\&\quad  B_{\rm ue}c + \tilde B_{m}c +\omega_{\vartheta,m,k}.
    \end{split} 
\end{align}
Now, the new random terms can be treated as additional noise components and incorporated into the error covariance matrices $\bm\Sigma_\tau$, and $\bm\Sigma_\vartheta$. Specifically, we augment both the diagonal entries and the entries corresponding to correlated measurements with a term %$\delta_m^2 c^2 / 3$.
$\delta_m^2 c^2$. The updated noise covariance matrices are then given by $\tilde{\bm \Sigma}_\tau = \bm \Sigma_\tau + \bm \Sigma_{\rm im}$ and $\tilde{\bm \Sigma}_\vartheta = \bm \Sigma_\vartheta + \bm \Sigma_{\rm im} $ where 
\begin{equation}
    \bm \Sigma_{\rm im} = %\tfrac{c^2}{3} 
    c^2\bm 1_{K\times K} \otimes \diag{[\delta_1^2,\ldots,\delta_M^2]^\top}.
\end{equation}
This formulation captures the full correlation of the imperfections across the bands while assuming independence across different BSs.
To also account for full correlation between \ac{TOA} and carrier-phase measurements, the composite covariance matrix is re-expressed as
\begin{equation}
    \tilde{\bm \Sigma}_{\rm ch} = \bm \Sigma_{\rm ch} + \bm 1_{2\times2}\otimes\bm \Sigma_{\rm im}.
\end{equation}
%The same update should be applied in the estimation process described in the next section. 
%
% I RELOCATED THIS TO A BIT LATER POINT IN TEXT
%

\vspace{-4mm}
\section{Proposed Multi-band CPP Estimators}
\label{section:sec4}
%To maintain a determined system of equations,
\textls[-2]{To determine the UE position with a practical estimation procedure, we divide the CPP problem into two stages. First, we obtain an \emph{initial estimate} of the UE coordinates via the delay measurements following the closed-form estimation approach in \cite{8494707} extended to multi-band measurements. Then, in the second stage, we consider both the delay and phase observations to establish the \emph{final estimate}. Given the initial coordinate estimate from the first stage, we linearize the observation model around it. %and define a new unknown update parameter.
We then compute the relaxed integer ambiguities using a weighted least squares (WLS) approach and subsequently resolve them in the integer domain using integer least squares techniques. With both the initial coordinate estimate and integer ambiguities determined, the UE position estimate is refined accordingly.}

\vspace{-3mm}
\subsection{First Stage: TDoA-based Initial Estimation}
For the \ac{TOA} measurements in \eqref{toa measurements}, there are well-known localization methods based on time difference of arrival (TDoA) combinations \cite{301830,1165089,259534,8494707}, as well as gradient-based techniques such as Gauss-Newton or gradient descent. Among these, the algorithm proposed by Amiri \emph{et al.} \cite{8494707}, inspired by the spherical intersection method \cite{mellen2003closed}, offers several advantages. First, it provides a closed-form solution, making it computationally efficient while eliminating the need for any hyperparameter tuning. Second, as noted in \cite{8494707}, it is performance-efficient and converges to the corresponding CRB -- a behavior that we also observe in our simulations. Finally, the algorithm requires only a modest number of measurements, i.e., a minimum of four in a three-dimensional setting.

While the original formulation in \cite{8494707} addresses or considers a classical single-band case, we extend it to the multi-band scenario. To this end, we use the measurement corresponding to the first band and the first BS as the reference\footnote{In general, the measurement with the lowest error variance should be chosen as the reference. However, since the error variances of different measurements are similar, we assume the first band and BS as the reference.} and then construct the $KM-1$ {TDoA} measurements. The remainder of the algorithm remains unchanged. For further details, the reader is referred to \cite{8494707} for presentation brevity.

\vspace{-2mm}
\subsection{Second Stage: Applying Carrier Phase Measurements}
% After finding an initial estimation of $\bm{s}$ from \eqref{eq:diff_},  denoted by $\bm{s}_0$, we can linearize the entire observation model around $\bm{s}_0$ using a first-order Taylor expansion.  Since the original observation model \eqref{eq:diff_} is already linear wrt. $B_{\rm ue},\bm{\varphi}_{\rm d}$, their initial values do not affect the linearization process.  Hence, we set ${\bm s}_0 = [\hat{\bm x}_{\rm ue}^\top,\bm{0}_{1\times (K+1)}]^\top $. The Taylor expansion of the observation model around ${\bm s}_0 $ yields
%In the first stage, we obtained an initial estimate of the UE position from equation \eqref{toa measurements}, denoted by $\hat {\bm{x}}_{{\rm ue},0}$. 
In the second stage, we refine the initial estimate -- denoted by $\hat {\bm{x}}_{{\rm ue},0}$ -- by incorporating also the carrier phase measurements. Since the observation model is nonlinear with respect to the UE position and cannot be solved analytically, we linearize it around $\hat{\bm{x}}_{{\rm ue},0}$ using a first-order Taylor expansion. The parameters $B_{\rm ue}$ and $\bm{\varphi}_{\rm d,ue}$ appear linearly in \eqref{eq:diff_}, and thus their initial values can be set arbitrarily;  for simplicity, they are initialized to zero. Accordingly, we define the initial estimate of the unknown vector $\bm s$ as ${\bm s}_0 = [\hat {\bm{x}}_{{\rm ue},0},\bm{0}_{1\times (K+1)}]^\top $, and the Taylor expansion of the observation model around ${\bm s}_0$ yields
\begin{equation}
\label{eq: linearized}
\bm{y} \simeq  \bm f(\bm{s}_0) + \bm A_{\bm f}(\bm{s}_0)\bm{\delta s} + \bm{B}\bm{E}\bm{z}_{\rm d} + \bm{\omega},
\end{equation}
where $\bm A_{\bm f}(\bm{s}_0) = \nabla_{\bm s}\bm f (\bm s)\big|_{\bm s_0} \in \mathbb{R}^{2KM\times {(K+{N_d}+1)}}$ is the Jacobian matrix evaluated at $\bm{s}_0$. This can be expressed as
\begin{equation}
\label{eq: gradient of s0}
\bm A_{\bm f}(\bm{s}_0)= 
\begin{bmatrix}
   {\bm{\tilde{U}}}^\top & c\times {\bm{1}_{KM\times 1}} & {\bm{0}_{KM\times K}} \\
   {\bm{\tilde{U}}}^\top & c\times {\bm{1}_{KM\times 1}} & %{\frac{1}{2\pi}}\times
   {\boldsymbol{\tilde{\Lambda}}}^\top
\end{bmatrix},
\end{equation}
where ${\boldsymbol{\tilde{\Lambda}}}= \text{diag}([\lambda_1,\cdots,\lambda_K]^\top)\otimes{\bf 1}_{ \rm{1\times M}}\in {\mathbb{R}}^{\rm K \times \rm{KM}}$, and $\bm{\tilde{U}}^\top$ is now calculated at the estimated coordinates $\hat{\bm x}_{\rm ue,0}$. 

Next, by moving all known terms to the left-hand side of \eqref{eq: linearized} and grouping the unknown variables into a single vector, we obtain the compact linear form
\begin{equation}
\label{eq: compact-linearized}
\tilde{\bm{y}} \simeq 
\begin{bmatrix}
   \bm A_{\bm f}(\bm{s}_0)& \bm{B}\bm{E}
\end{bmatrix}
\begin{bmatrix}
   \bm{\delta s} \\ \bm{z}_{\rm d}
\end{bmatrix}
 + \bm{\omega},
\end{equation}
where $\tilde{\bm{y}} = {\bm{y}} - \bm{f}(\bm{s}_0)$. 
%Let $\bm{\Sigma}_{\rm ch}$ denote the covariance matrix of the noise vector $\bm{\omega}$. 
We now apply the WLS method to obtain the relaxed estimate $\hat{\bm{z}}_{\rm d,rlx} \in \mathbb{R}^{K(M-1)\times 1}$ of $\bm{z}_{\rm d}$. To recover the integer estimate of $\bm{z}_{\rm d}$, we follow the same procedure outlined in problem \eqref{integer-prog}, where the matrix ${\bm{S}}$ is replaced here with the error covariance of the estimation method, denoted by $\widehat{\bm{S}}$. %Accordingly, the transformation becomes ${\bm{r}} = \widehat{\bm{S}}^{-1/2} \bm{z}_{\rm d,rlx}$.
 Here, $\widehat{\bm{S}}$ represents the error covariance matrix associated with the WLS estimate of $\bm{z}_{\rm d}$ from the observation model \eqref{eq: compact-linearized}, and it corresponds to the lower-right $K(M-1) \times K(M-1)$ block of the full WLS covariance matrix $\widehat{\bm{C}}$, given by\footnote{Alternatively, one can also use $\widehat{\bm{S}} = \bm D \hat{\bm\Sigma}_{\rm rlx}\bm D$, where $\hat{\bm\Sigma}_{\rm rlx}$ is computed at $\hat {\bm{x}}_{{\rm ue},0}$. However, the covariance matrix derived from \eqref{error-covariance} is preferred as it represents the actual error covariance of $\hat{\bm{z}}_{\rm d,rlx}.$}
\begin{equation}
\label{error-covariance}
    \widehat{\bm C} = \Bigl[\Bigl(\begin{bmatrix}
    \bm A_{\bm f}(\bm{s}_0)\ \bm{B}\bm{E}
    \end{bmatrix}^ \top
    \bm \Sigma_{\rm ch}^{-1} 
    \begin{bmatrix}
    \bm A_{\bm f}(\bm{s}_0)\ \bm{B}\bm{E}
    \end{bmatrix}\Bigr)^{-1} 
    \Bigr].
\end{equation}
It is worth noting that any well-known integer programming method, such as the Least-squares AMBiguity Decorrelation Adjustment (LAMBDA) approach \cite{teunissen1993least}, can be employed to solve the related integer problem. % in \eqref{integer-prog}. 
Once the integer estimate $\hat{\bm z}_{\rm d}$ is obtained, we return to \eqref{eq: compact-linearized}, move the known terms to the left-hand side, and solve for the remaining unknown, which is the update term $\bm{\delta s}$. The final WLS problem then takes the following form
\begin{equation}
\label{eq: 2compact-linearized}
\tilde{\tilde{\bm{y}}} \simeq 
\bm A_{\bm f}(\bm{s}_0) \bm{\delta s} + \bm{\omega},
\end{equation}
where $\tilde{\tilde{\bm{y}}} = \tilde{\bm{y}} - \bm{B}\bm{E}\hat{\bm{z}}_{\rm d} $. After estimating $\widehat{\bm{\delta s}}$, the solution is updated as $\hat{\bm{s}} = \bm{s}_0 + \widehat{\bm{\delta s}}$.

To further improve the estimation accuracy, the second stage of the algorithm can be repeated iteratively. At the end of each iteration, the updated solution $\hat{\bm{s}}$ is used as the new linearization point $\bm{s}_0$ for the next iteration. The final estimate is taken as the value of $\hat{\bm{s}}$ obtained in the last iteration. As demonstrated in the numerical results, only two iterations of the second stage are typically sufficient to achieve high estimation accuracy, although a more general exit or stopping criterion can also be considered. Overall, as demonstrated through the numerical results in Section V, the algorithm converges efficiently to the mixed-integer bound under a broad range of system parameters, and can thus be called an efficient estimator. % sufficiently favorable system conditions.

\vspace{-3mm}
\subsection{Search-based Refinement of the First Stage Solution}
Since the maximum likelihood (ML) cost function of the observation model in \eqref{eq:diff_} is non-convex and may contain numerous local minima, obtaining a sufficiently accurate solution in the first stage of the algorithm is crucial. However, the accuracy achievable in the first stage is fundamentally limited by the delay-only PEB. Therefore, as we will observe through the numerical results, under narrowband configurations, even efficient delay-only estimators may fail to provide an adequately accurate initial solution for the CPP problem.

To address this challenge, we propose to estimate the delay-only FIM at the obtained initial solution, $\bm s_0$, and use the corresponding error covariance matrix to define a search region around $\bm  s_0 $. Within this region, a set of candidate points is generated, either randomly or deterministically. The second-stage procedure is then applied to each of these points, and the one yielding the lowest ML cost (considering both delay and phase measurements) is selected. Let $\mathcal{C}_{\bm s}$ with $|\mathcal{C}_{\bm s}|=N_s$ denote the set of all initial candidate points, and $\mathcal{C}_{\hat{\bm s}}$ is the set of the corresponding outputs of the second stage. The final solution is obtained by solving the following discrete optimization problem of the form
\begin{equation}
\begin{split}
     n^\star =   \underset{n\in \{ 1, \cdots,N_s \} }{\arg \min} \ & \Big(\bm{y} - {\bm f}(\hat{\bm s}^{(n)}) - \bm{B}\bm{E}\hat{\bm{z}}_{\rm d}^{(n)} \Big)^\top \bm\Sigma_{\rm ch}^{-1}\\&
     \times\Big(\bm{y} - {\bm f}(\hat{\bm s}^{(n)}) - \bm{B}\bm{E}\hat{\bm{z}}_{\rm d}^{(n)} \Big)
\end{split}
   \label{ML cost}
\end{equation}
 where $\hat{\bm s}^{(n)}$ and $\hat{\bm{z}}_{\rm d}^{(n)}$ are the estimated parameters when the $n$-th initialization point is used for the second stage. %corresponding to the selected ${\bm s}_0$.     
The final estimate is $\bm s^\star = \bm s^{(n^\star)}$ with ${n^\star}$ obtained from \eqref{ML cost}.
Increasing the number of candidate points enhances the likelihood of approaching the MICRB, as more points are likely to fall within the attraction region of the final optimal solution. However, this comes at the cost of higher computational complexity. %, which will be addressed in future works. 

\textls[-3]{The overall proposed approach is summarized in Algorithm~\ref{alg:algorithm1}.}

\vspace{-4mm}
\subsection{Complexity Analysis}
The first stage of the algorithm, which finds the TDoA solution in closed-form based on the multi-band extension of \cite{8494707}, has a computational complexity on the order of $\mathcal{O}((KM - 1)^3)$. 
In the second stage, solving the real-domain WLS problems has a complexity of $\mathcal{O}(K_d^3 + KMK_d^2 + K^2M^2K_d)$, where $K_d = K + N_d + 1$. 
For the integer ambiguity resolution, the LAMBDA method has an overall complexity of $\mathcal{O}((K(M - 1))^3 + \kappa^{K(M - 1)})$, where $1 < \kappa \leq 2$. The first term corresponds to the decorrelation (reduction) stage, and the second term corresponds to the search stage \cite{teunissen1993least}.  However, as shown in \cite{chang2005mlambda}, the Modified LAMBDA (MLAMBDA) method significantly improves efficiency in both stages by optimizing the search parameter $\kappa$.
Finally, note that in the second stage, the overall complexity scales linearly with the number of search points, $N_s$, and number of iterations, $N_{\rm iter}$. 

Overall, it is fair to conclude that there is notable computing complexity involved. Hence, developing reduced-complexity yet efficient estimators is an important topic for future research.

\begin{algorithm}[!t]
\caption{Proposed Two-Stage TDoA-CPP Algorithm}
\small
\label{alg:algorithm1}
\begin{algorithmic}[1]
\State  Choose $N_{\rm iter} \geq 1$, $\epsilon$, $N_{\rm s}$  , and  set $t = 1$.
\State Estimate the initial solution $\bm{s}_0$ based on the \ac{TOA} observations in \eqref{toa measurements}, using the multi-band extension of the algorithm in \cite{8494707}.
\State Find $\bm{\Sigma}_{\text{delay}}(\bm{s}_0)$ based on \eqref{j_delay}.
\State Pick $N_{\rm s}$ random or deterministic points from the area defined by $\epsilon\bm{\Sigma}_{\text{delay}}(\bm{s}_0)$ around $\bm{s}_0$ and build the set $\mathcal{C}_{\bm s}$.

\For{each  $\bm s \in \mathcal{C}_{\bm s} $}
\State Set $\bm{s}_0 \leftarrow {\bm{s}}$.
\Repeat
\State 
    Linearize the full set of observations around  $\bm{s}_0$  to \Statex \hspace{1cm} obtain  the model in \eqref{eq: compact-linearized}.
\State  Estimate the relaxed real-valued solution $\hat{\bm{z}}_{\rm d,rlx}$ \Statex \hspace{1cm} by  applying WLS to \eqref{eq: compact-linearized}.
\State Estimate the integer ambiguities $\hat{\bm{z}}_{\rm d}$ by solving  the \Statex \hspace{1cm} ILS  problem in \eqref{integer-prog}, using the error covariance \Statex \hspace{1cm} $\widehat{\bm{S}}$  from \eqref{error-covariance}.
\State  Substitute $\hat{\bm{z}}_{\rm d}$ into \eqref{eq: compact-linearized} to obtain the model in  \eqref{eq: 2compact-linearized}, \Statex \hspace{1cm} then estimate  $\widehat{\bm{\delta s}}$.
\State  Update the solution: $\hat{\bm{s}} = \bm{s}_0 + \widehat{\bm{\delta s}}$.
\State  Set $\bm{s}_0 \leftarrow \hat{\bm{s}}$.
\State  Increment $t \leftarrow t + 1$.
\Until{$t > N_{\rm iter}$}
\State Buffer $\hat{\bm{s}}$ , $\hat{\bm z}_{\rm d}$.
\EndFor
\State Solve the problem \eqref{ML cost} to obtain $\bm s^\star$ and $\bm x_{\rm ue}^\star$.
\end{algorithmic}
\end{algorithm}

%\section{Key Influences on Positioning Performance}
\vspace{-2mm}
\section{Nonuniform BS-Band Assignment}
\label{section:sec5}
So far, we have assumed that all BSs communicate using the same set of frequency bands. In this section, for improved deployment flexibility, we generalize the proposed algorithm to scenarios with nonuniform BS-band assignments, where different BSs may transmit at different subsets of bands. We also consider an extreme case in which BSs operate at completely disjoints sets of bands, thereby highlighting the importance of providing a band-independent phase offset.  

To this end, to extend the proposed algorithm to nonuniform BS-band assignments, we simply rearrange the observation model in \eqref{eq:obs_model} by stacking and grouping the observations corresponding to each band.
In scenarios with band-dependent phase offsets ($\varphi_{{\rm ue},k}$), at least two phase measurements per band (i.e., two BSs have to transmit on the same band) are required to make the data exploitable for the CPP; otherwise, the terms $z_{m,k}$ and $\varphi_{{\rm ue},k}$ become indistinguishable. In contrast, when the phase offset is band-independent, the algorithm can be generalized to completely disjoint BS-band assignments. %without such constraints.
In the following subsections, we first address the case of nonuniform BS-band assignment under band-dependent phase offsets, and then consider the case of band-independent phase offsets.

\vspace{-3mm}
\subsection{Band-dependent Phase Offsets}
\textls[-4]{As noted above, for each band, we group the measurements and rearrange the observations. The first stage of the algorithm remains the same, applying TDoA-based localization to the new set of observations. In the second stage, the matrices used in the observation model \eqref{eq: linearized} must be updated accordingly. To this end, we first construct the matrices $\bm A_{\bm f}(\bm{s}_0)$ and $\bm B$ assuming a full assignment of each BS to all existing bands (i.e., all $KM$ possible combinations). We then remove the rows and columns corresponding to non-existing assignments. To construct $\bm z_{\rm d}$, we designate the first BS assigned to each specific band as the reference BS for that band and define the differential integer ambiguities accordingly. Based on this, we can derive and express the updated matrix $\bm E$. The same modification applies when deriving the performance bounds. The exact mathematical description is omitted for presentation brevity.}

\vspace{-3mm}
\subsection{Band-independent Phase Offsets}
{{A band-independent phase offset is feasible and observed in practice when the responses of the transmitter and receiver chains (i.e., amplifiers, mixers, digital-to-analog/analog-to-digital converters, etc.)  at the different bands are calibrated, that is to say, measured before using the system and compensated digitally during the subsequent system operation \cite{9555252}. We note that in any case, a single shared oscillator has to be used in the transmitter and also in the receiver, since this is a prerequisite for having a band-independent clock bias, $B_{\rm ue}$ -- an assumption that has been made throughout the article and work. 
%A band-independent phase offset is observed in practice when both the transmitter and receiver employ a single RF chain, including amplifiers, oscillators, and digital-to-analog/analog-to-digital conversion, for the entire multi-band aggregated bandwidth \cite{9555252}. 

\textls[-4]{The band-independent phase offset scenario enables the CPP framework to handle a wider range of BS-band assignment types, including fully disjoint assignments, which are particularly useful under bandwidth resource constraints.}} We thus next first formulate the band-independent phase offset case for the fully uniform BS-band assignment, and then apply the same update procedure from the previous subsection to extend it to nonuniform assignments and completely disjoint sets of bands. Before proceeding, we note that, for the band-independent phase offset case, we use $\varphi_{\rm ue}$ instead of $\varphi_{{\rm ue},k}$, and the new set of real-valued unknowns becomes $\bm{\bar {s}}=\left[{{\bm{x}}}_{{\rm ue}}^\top, {B}_{\rm ue}, {\varphi}_{\rm d} \right]^\top\in \mathbb{R}^{({N_d}+2)\times 1}$. In the fully uniform assignment case, ${\varphi}_{\rm d} = {\varphi_{\rm ue}} + z_{1,1}$ when choosing the first unknown integer ambiguity as the reference, again without lost of generality.}

Since the delay measurements are intact, the first stage of the algorithm remains the same. However, due to the new definition of $\bm {\bar s}$, the linearized observation~\eqref{eq: linearized} takes now the following form
\begin{equation}
\label{eq: linearized phi-ind}
\bm{y} \simeq  \ {\bm {\bar f}}({\bm {\bar s}}_0) + \overline{\bm  A}_{\bm f}(\bm{\bar s}_0)\overline{\bm{\delta s}} + \bm{B}\overline{\bm  E}{\bm{\bar z}}_{\rm d} + \bm{\omega}.
\end{equation}
Here, ${\bm {\bar f}}({\bm {\bar s}}) = \tilde{\bm f}(\tilde{\bm s}) +{\rm diag}(\bm{B}) {\varphi}_{\rm d}$, and the Jacobian matrix evaluated at $\bm{\bar s}_0$ is $\overline{\bm  A}_{\bm f}(\bm{\bar s}_0) = \nabla_{\bm {\bar s}}{\bm {\bar f}}({\bm {\bar s}}) \big|_{{\bm \bar s}_0} \in \mathbb{R}^{2KM\times ({N_d}+2)}$. This can now be expressed as
\begin{equation}
\label{eq: gradient of bar s0} 
\overline{\bm  A}_{\bm f}(\bm{\bar s}_0) = 
\begin{bmatrix}
   {\bm{\tilde{U}}}^\top & c\times {\bm{1}_{KM\times 1}} & {\bm{0}_{KM\times 1}} \\
   {\bm{\tilde{U}}}^\top & c\times {\bm{1}_{KM\times 1}} & %\frac{1}{2\pi}\times
   {\boldsymbol{\bar{\lambda}}}^\top
\end{bmatrix},
\end{equation}
where ${\bm{\bar{\lambda}}}= [\lambda_1,\cdots,\lambda_K]^\top\otimes{\bf 1}_{ \rm{1\times M}}\in {\mathbb{R}}^{\rm 1 \times \rm{KM}}$.
The differential integer ambiguity is $\bm{\bar z}_{\rm d} = \overline{\bm  D} \bm z$, where the differentiator matrix $\overline{\bm  D}$ has the form $\overline{\bm  D} =  \begin{bmatrix}-\bm{1}_{(KM-1)\times 1} & \bm{I}_{(KM-1)}\end{bmatrix} \in \mathbb{R}^{(KM-1)\times KM}$, and the matrix $\overline{\bm{ E}} = \begin{bmatrix}\bm{0}_{(KM-1)\times 1 }, \bm{I}_{KM-1}\end{bmatrix}^\top \in \mathbb{R}^{KM \times (KM-1)}$.

The remaining steps proceed as described in the second stage. To extend the algorithm to the nonuniform assignment case, or to completely disjoint bands, which represent a specific case of nonuniform assignment, we can select any existing integer ambiguity as the reference to define $\varphi_{\rm d}$ and construct $\bm{\bar z}_{\rm d}$. The matrices $\overline{\bm{D}}$ and $\overline{\bm{E}}$ are then trivially updated accordingly. Additionally, for $\overline{\bm{A}}_{\bm{s}_0}$ and ${\bm{B}}$, we remove the rows and columns corresponding to measurements that no longer exist.

}

\vspace{-3mm}
\section{Numerical Results and Insights}
\label{section:sec6}
%\textcolor{blue}{Note that the Slepian-Bangs formula provides a fundamental link between the FIM and performance bounds in parameter estimation, particularly in signal processing and localization. FIM analysis quantifies the amount of information a signal carries about an unknown parameter, enabling the derivation of lower bounds on estimation accuracy, such as the Cramér-Rao bound [REF required]}

%In this part we 
We next investigate and evaluate the derived multi-band CPP bounds in a concrete network deployment context. {\color{black}{ Furthermore, we assess the accuracy and robustness of the proposed estimators in this setting and demonstrate the tightness of the derived bound, identifying the conditions under which the estimator attains it.}}
%proposed mixed integer bound for different setups.
We consider an example physical scenario where the UE is fixed at the location $\bm{x}_{{\rm{ue}}} = [0\, 0]^{\top}$, while $M$ BSs or other network nodes/TRPs are located in the surrounding 2D space with their positions $\bm{x}_{{\rm{bs}}, m}$ following a normal distribution as $\bm{x}_{{\rm{bs}}, m} \sim \mathcal{N}(\bm{0}, (0.1 \, \mathrm{km})^2 \bm{I}_{2 \times 2})$. 
%
%{\color{blue}The realization used in simulations is shown in Fig.~\ref{fig:BSUEmap}}. 
% Commented out to save space
\textls[-2]{The following default system parameters are utilized, unless otherwise stated: subcarrier spacing $\Delta_{{ f},k} = 30\,\mathrm{kHz}$, $N_k = 612$ subcarriers (51 NR resource blocks, per band), noise power spectral density of $-174 \, \mathrm{dBm}/\mathrm{Hz}$, a receiver noise figure of $13 \, \mathrm{dB}$, and $M = 6$ BSs.  We model the magnitude of the channel gain as $\rho_{m,k} = {\lambda_{\rm ref}}/{\left(4 \pi \|\bm{x}_{{\rm{bs}}, m} - \bm{x}_{{\rm{ue}}}\|\right)}$ with $\lambda_{\rm ref} = 3 \, \rm{cm}$. Therefore, we assume that transceiver array gain essentially compensates the dependence of the path loss factor with respect to a reference frequency of $10\,\mathrm{GHz}$. %For reference, 
The default per-band transmit power is $P_{{\rm tx},{k}} = 0 \, \mathrm{dBm}$.
As the default carrier frequencies at different bands, we use $f_c = 3.5  \,\rm GHz$ (FR1), $f_c = 28 \,\rm  GHz$ (FR2), and $f_c = 12 \,\rm GHz$ (FR3), unless otherwise stated. The number of iterations for the second stage of Algorithm~\ref{alg:algorithm1} is set to $N_{\rm iter} = 2$ and the default number of search points is set to $N_s = 1$.}

% \begin{figure}[!t]
% 		\centering
% 		\vspace{-0mm}
%         \includegraphics[width=7.6cm,trim={0cm 0.2cm 1cm 0.5cm}]{BS_UE_map.eps}
%         \vspace{-1.5mm}
% 		\caption{2D map of BSs and the UE. The numbers near BSs indicate their selection order when varying the number of BSs. }
%         \vspace{1mm}
% 		\label{fig:BSUEmap}
% \end{figure}

For assessing and comparing the impacts of the different parameters, we use the PEB defined as ${\rm PEB} = \sqrt{{\rm tr}  (\bm\Sigma_\star (\bm x _{\rm ue}))}$ for all the mixed-integer, known-integer, and delay-only bounds, while the evaluation  metric for the estimator is the root-mean-square error (RMSE) expressed as:
\begin{equation}
   \mathrm{RMSE} = \sqrt{\frac{\sum_{i=1}^{N_{mc}}\|\hat{\bm x}_{{\rm ue},i}-\bm x_{\rm ue}\|^2}{N_{mc}}} \ , 
\end{equation}
The proposed MICRB given in (\ref{j_mixed}) and the RMSE are computed over $N_{\rm mc}=1\,000$ Monte-Carlo samples. 

\vspace{-1.5mm}
\subsection{Carrier Frequency and Multi-band Measurements}

{\color{black}
{{\color{black}We start by illustrating the impact of carrier frequency while also comparing the single-band and dual-band scenarios. The results are shown in Fig.~\ref{fig:PEBvsfc}. In the single-band case, $f_{c,1}$ varies, whereas in the dual-band case $f_{c,1}$ is fixed and $f_{c,2}$ varies. 
The delay-only PEB ($\rm{PEB}_{\rm{delay}}$) appears unaffected by the carrier frequency, as it primarily depends on SNR and bandwidth rather than the frequency (assuming a constant SNR across bands). In contrast, the known-integer PEB ($\rm{PEB}_{\rm{known}}$) -- plotted only for the dual-band case for the readability of figure -- %, where the integer ambiguities are assumed to be correctly resolved,
improves with increasing carrier frequency since shorter wavelengths reduce the power of the measurement noise in (\ref{phase measurements}). In mathematical terms,  $\rm{PEB}_{\rm{known}}$ is governed by $\bm \Sigma_\vartheta$, which improves as $\lambda$ decreases. For higher values of $f_{c,2}$, the influence of $f_{c,1}$ disappears, and both dual-band cases coincide.  }

Furthermore, the behavior of the proposed mixed-integer PEBs ($\rm{PEB}_{\rm{mi}}$) reveals more interesting insights. In the single-band case, the $\rm{PEB}_{\rm mi}$ curves converge to $\rm{PEB}_{\rm known}$ for carrier frequencies below $f_{c,1} = 9\,\rm{GHz}$ (with doubled TX power) and $f_{c,1} = 5\,\rm{GHz}$ (default TX power). This means that the reduced-state ambiguities are fully resolved below these frequency thresholds. At higher frequencies, however, the curves diverge since shorter wavelengths lower the probability of correct ambiguity resolution. In other words, the integer program \eqref{integer-prog} becomes harder to solve correctly at higher frequencies.
In the dual-band case, unlike single-band scenario, no divergence is observed, and $\rm{PEB}_{\rm mi}$ coincides with $\rm{PEB}_{\rm known}$ across all $f_{c,2}$ values. This improved performance stems from the diversity provided by measurements whose ambiguities are multiples of different wavelengths,which enables more reliable ambiguity resolution. Even when $f_{c,1} = 28 \, \rm{GHz}$ and   $f_{c,2}>100\, \rm GHz$ , we still have $\rm{PEB}_{\rm{mi}}=\rm{PEB}_{\rm{known}}$, whereas in the single-band cases, the convergence does not occur for either $f_{c,1} = 28 \, \rm{GHz}$ or $f_{c,1}\geq100\, \rm GHz$.  }
}

\begin{figure}[!t]
		\centering
		\vspace{-1mm}
        \includegraphics[width=8.2cm,trim={0cm 0.2cm 1cm 0.5cm}]{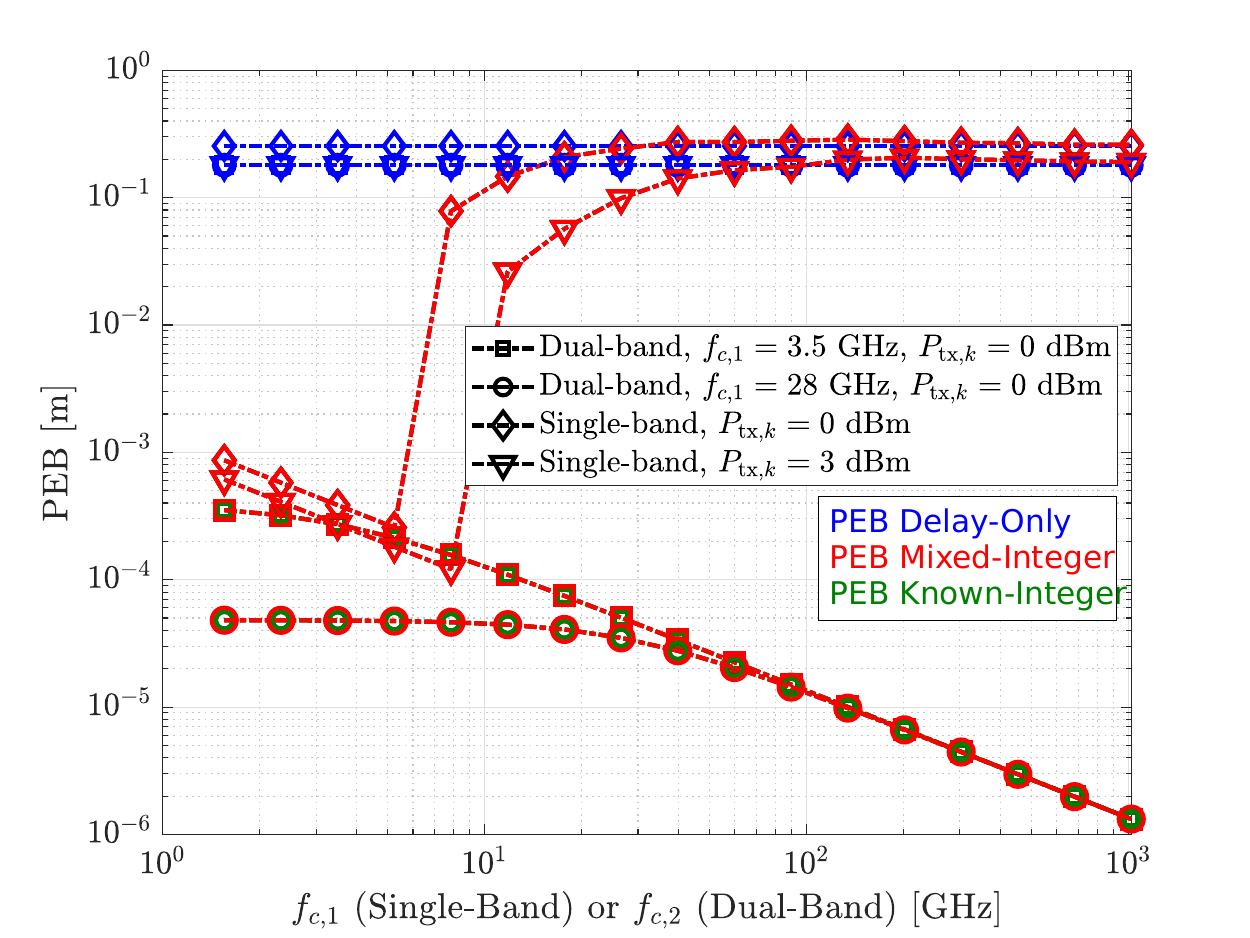}
        \vspace{-0.5mm}
		\caption{Impact of the carrier frequency on the different PEBs. For the single-band case, a doubled $P_{\rm tx}$ case is also shown for fair comparison.}
        \vspace{-4mm}
		\label{fig:PEBvsfc}
\end{figure}

\textls[-1]{Next, Fig.~\ref{fig:PEBvsK} illustrates the combined effect of carrier frequency and the number of measurement bands, $K$. To fairly assess the impact of the number of bands, we %increase $K$ while maintaining 
maintain constant sum-power of $K \times P_{{\rm tx},{k}} = 0\,{\rm dBm}$ for any given $K$. For $K=1$, all curves correspond to $f_{c,1} = 3.5 \, \mathrm{GHz}$, while when $K$ increases from $2$ to $10$, we include sequentially the following carrier frequencies in the different bands: 
FR1: $\left[3.6\!:\!0.2\!:\!5.2\right] \, \mathrm{GHz}$,
FR2: $\left[24\!:\!1\!:\!32\right] \, \mathrm{GHz}$, and
FR3: $\left[8\!:\!0.5\!:\!12\right] \, \mathrm{GHz}$. Thus overall, FR1 adopts intra-band CA, while the other cases correspond to inter-band CA.}
As shown in Fig.~\ref{fig:PEBvsK}, $\text{PEB}_{\text{delay}}$ remains constant as the number of bands $K$ increases. This is because the benefit of additional measurements is balanced by the reduced transmit power per band, and the new measurements do not provide a different type of information;  only the noise realizations differ across bands. In contrast, $\text{PEB}_{\text{mi}}$ and $\text{PEB}_{\text{known}}$ improve for two reasons. First, having more bands introduces additional measurements that contain new information about the location, as the integer ambiguities differ across bands. This effect is especially noticeable when increasing from a single-band to dual-band. Second, the higher carrier frequencies of the added bands tighten the bound, as previously discussed in Fig.~\ref{fig:PEBvsfc}. For $K\geq2$, $\rm{PEB}_{\rm{mi}}$ is completely coincided with $\rm{PEB}_{\rm{known}}$.
%, while the mutual differences for larger values of $K$ are primarily influenced by the carrier frequency of the newly added bands.

\begin{figure}[!t]
		\centering
		\vspace{-1mm}
        \includegraphics[width=8.2cm,trim={0cm 0.2cm 1cm 0.5cm}]{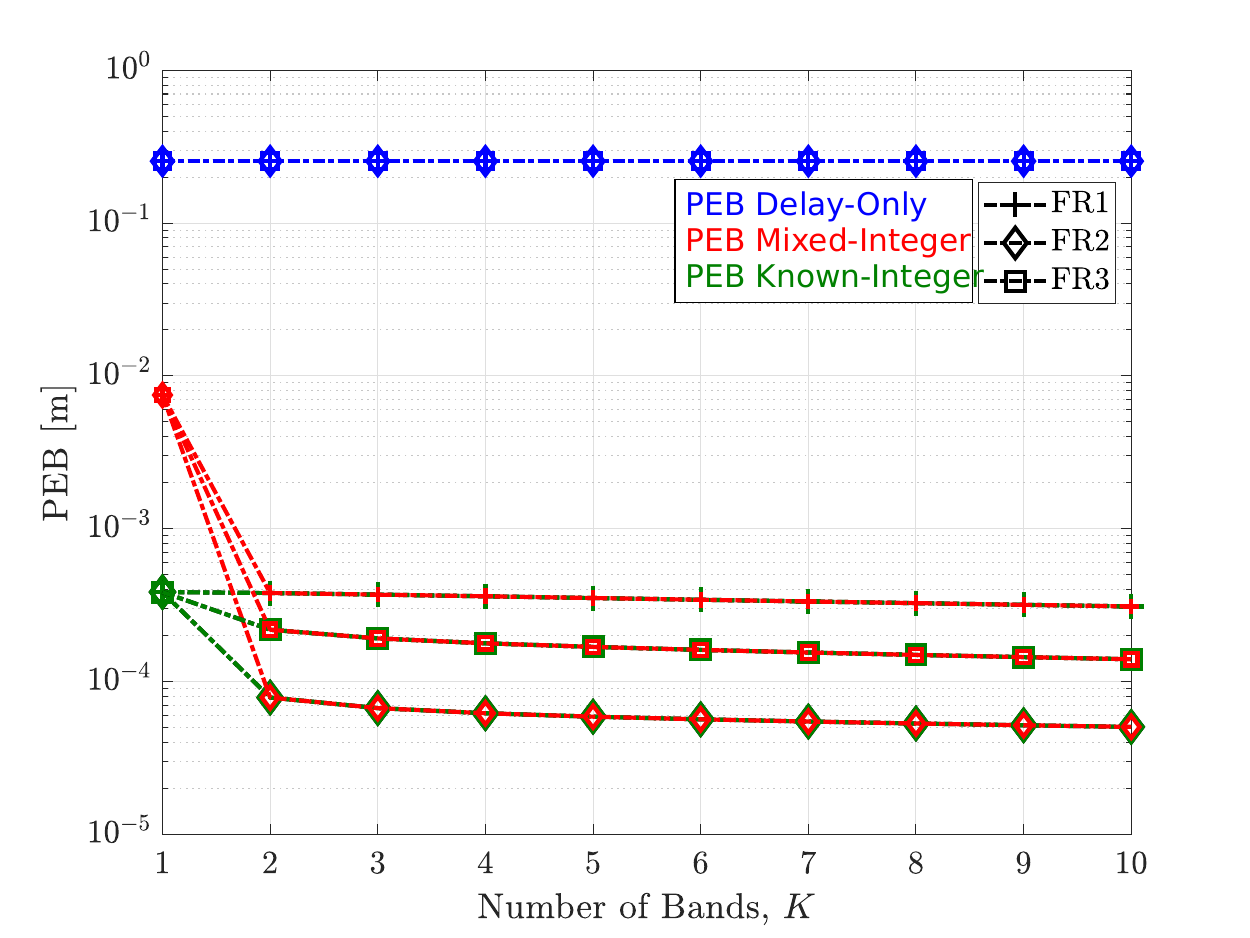}
		\vspace{-0.5mm}
        \caption{PEB vs. the number of measurement frequencies, $K$. %, with  increasing $f_{c,k}$ of the added bands. 
        In all cases, the first carrier frequency ($f_{c,1}$) correspond to 3.5\,GHz, while the additional frequencies follow the frequency ranges {indicated in the text.}  }
        \vspace{-3mm}
		\label{fig:PEBvsK}
\end{figure}

\vspace{-1.0mm}
%\subsection{Tightness Evaluation of the Proposed Multi-band Bound} 
\subsection{Comparison of the Proposed Multi-band Bound with Existing Single-band Bounds }
\vspace{-0.5mm}
To verify the %tightness 
optimality of the proposed multi-band mixed-integer bound, we provide a comparison between the single-band bounds proposed in \cite{10437902}, their fusion across bands, and the multi-band bounds in our work. To obtain the fusion of separate single-band cases, we sum up the effective FIM matrices\footnote{The effective part of the FIM matrix is obtained by computing the Schur complement of the FIM block associated with the phase offset.} of the position and clock bias parameters (i.e., the inverse of the upper-left $(N_d + 1)\times(N_d + 1)$ block of the full CRB matrix) of each band and compute the corresponding covariance matrices. This approach is followed because each band has distinct phase offsets as unknown parameters, so we focus on the effective FIM for the common parameters across all bands.

\textls[-4]{As shown in Fig.~\ref{fig:singleFusion}, the  $\rm{PEB}_{\rm{known}}$ and  $\rm{PEB}_{\rm{delay}}$ bounds are identical in both the multi-band and the band-fusion cases, since different bands provide linear measurements with independent noise and real-valued phase offsets which are solved independently for each band. Therefore, the multi-band {FIM} is equal to the sum of the individual FIMs. %However, for  $\rm{PEB}_{\rm{mi}}$, since the integer ambiguities are interrelated across the bands, the proposed multi-band bound combines the information coherently, enabling better ambiguity resolution and yielding a tighter bound than the band-fusion case. 
However, the behavior differs for $\rm{PEB}_{\rm{mi}}$.
In this case, the multi-band $\rm{PEB}_{\rm{mi}}$ formulation implicitly assumes that all integer ambiguities are resolved jointly, while the band-fusion $\rm{PEB}_{\rm{mi}}$ formulation assumes they are solved independently within each band. The observed improvement in the multi-band $\rm{PEB}_{\rm{mi}}$ therefore reflects the benefits of jointly estimating interrelated integer parameters (an effect absent in $\rm{PEB}_{\rm{known}}$ and $\rm{PEB}_{\rm{delay}}$, which  involve only real parameters). As a result, the joint resolution of two interrelated sets of integer ambiguities provides more reliable ambiguity estimation and yields faster convergence in the multi-band ${\rm PEB}_{\rm mi}$ compared to the band-fusion case.}

\textls[-4]{The effect of the number of BSs, $M$, is also assessed and highlighted in Fig.~\ref{fig:singleFusion}. While increasing $M$ generally improves performance, the improvement is not strictly linear due to the involved randomness in {BS} placement (additional BSs are dropped randomly in the area). Importantly, the multi-band mixed-integer PEB reaches convergence with fewer BSs than the other cases.}

%Increasing the number of base stations enhances the number of observations, leading to improved PEBs.

%\subsection{Comparison with Single-Band Bounds \cite{10437902}} This figure presents a fair comparison with the proposed bounds in \cite{10437902} for single-band cases. To achieve this, we sum the FIM matrices of different single-band cases and compute the corresponding covariance matrix. Since different bands have distinct phase offsets as unknown parameters, we consider only the effective part of the FIM, which corresponds to the user coordinates and clock bias—parameters common across all bands\footnote{The effective part of the FIM matrix is obtained by computing the Schur complement of the FIM block associated with the phase offset.}.
%As shown in Fig. \ref{fig:singleFusion}, the $\text{PEB}{\text {known}}$ and $\text{PEB}{\text {delay}}$ bounds are identical in both the  multi-band and single-band fusion cases. This is because, in these cases, different bands merely provide additional measurements with independent noise, and the  multi-band bound simply accumulates the available information while treating the band-dependent phase offset as a nuisance parameter. However, for $\text{PEB}_{\text {mi}}$, the proposed  multi-band bound combines information coherently, enabling significantly better resolution of integer ambiguities compared to the naive combination of single-band measurements.
\begin{figure}[!t]
		\centering
		\vspace{-1mm}
        \includegraphics[width=8.2cm,trim={0cm 0.2cm 1cm 0.5cm}]{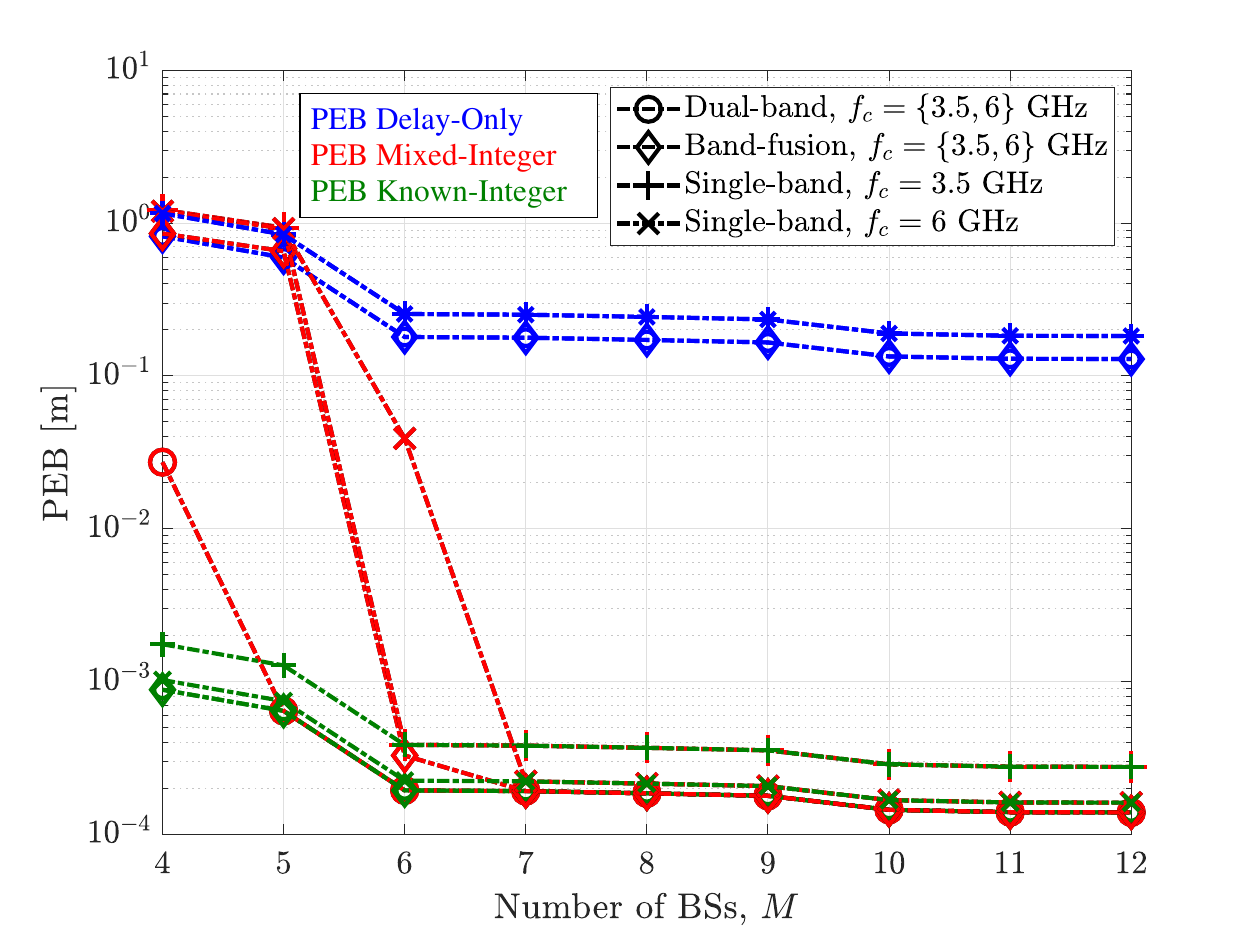}
		\vspace{-0.0mm}
        \caption{Comparison between the  multi-band and single-band fusion approaches while also varying the number of base stations, $M$.}
        \vspace{-4mm}
		\label{fig:singleFusion}
\end{figure}

{\color{black}
\subsection{Performance Evaluation of the Proposed Estimator}
\subsubsection{Search-based Optimization}
\textls[-12]{To evaluate the performance of Algorithm~\ref{alg:algorithm1}, we examine its RMSE curves under different bandwidths and varying numbers of candidate points ($N_s$), which are used as initial solutions after the algorithm's first stage. These candidate points are sampled randomly from a Gaussian distribution centered at the delay-only estimate, $\bm{s}_0$, with covariance $\bm{\Sigma}_{\rm{delay}}(\bm{s}_0)$. This choice ensures that the selected points lie within a sufficiently reliable confidence region and are close enough to each other. The corresponding $\text{PEB}_{\text{delay}}$ and $\text{PEB}_{\text{known}}$ bounds are used as benchmarks. Fig.~\ref{fig:searchBW} presents the results for a dual-band scenario involving FR1 and FR3 aggregation.}

\textls[-10]{As shown, the RMSE curves of the algorithm's first stage (delay-only) converge to the $\text{PEB}_{\text{delay}}$ starting from $2 \, \rm{MHz}$ bandwidth. However, in the case of $N_{s} = 1$, which corresponds to a scenario without a search process, the narrowband regime (less than $13.5 \, \rm{MHz}$) reveals that even the $\text{PEB}_{\text{delay}}$ accuracy is insufficient for full convergence in the second stage. Moreover, within this bandwidth range, large integer errors in several Monte-Carlo trials cause the final RMSE to exceed that of the delay-only stage. Nevertheless, the case of $N_{s} = 1$ remains practical for CPP in a wide range of applications, as the bandwidths available in 5G NR and future 6G networks typically exceed $13.5 \,\text{MHz}$.   As the number of search points increases to $N_{s} = 1000$, the convergence point of the algorithm improves from $W = 13.5 \,\text{MHz}$ to $W = 2.65 \,\text{MHz}$, making it more suitable for IoT applications. For $N_{s} = 10^5$,  full convergence is achieved across the entire bandwidth range, albeit at the cost of higher complexity. Increasing the number of random points raises their density within a fixed region and also expands the overall area they may occupy. These factors together make it more likely that some samples fall into the attraction region of the final optimal solution, thereby ensuring convergence to the multi-band MICRB.}

\begin{figure}[!t]
		\centering
		\vspace{-1mm}
        \includegraphics[width=8.2cm,trim={0cm 0.2cm 1cm 0.5cm}]{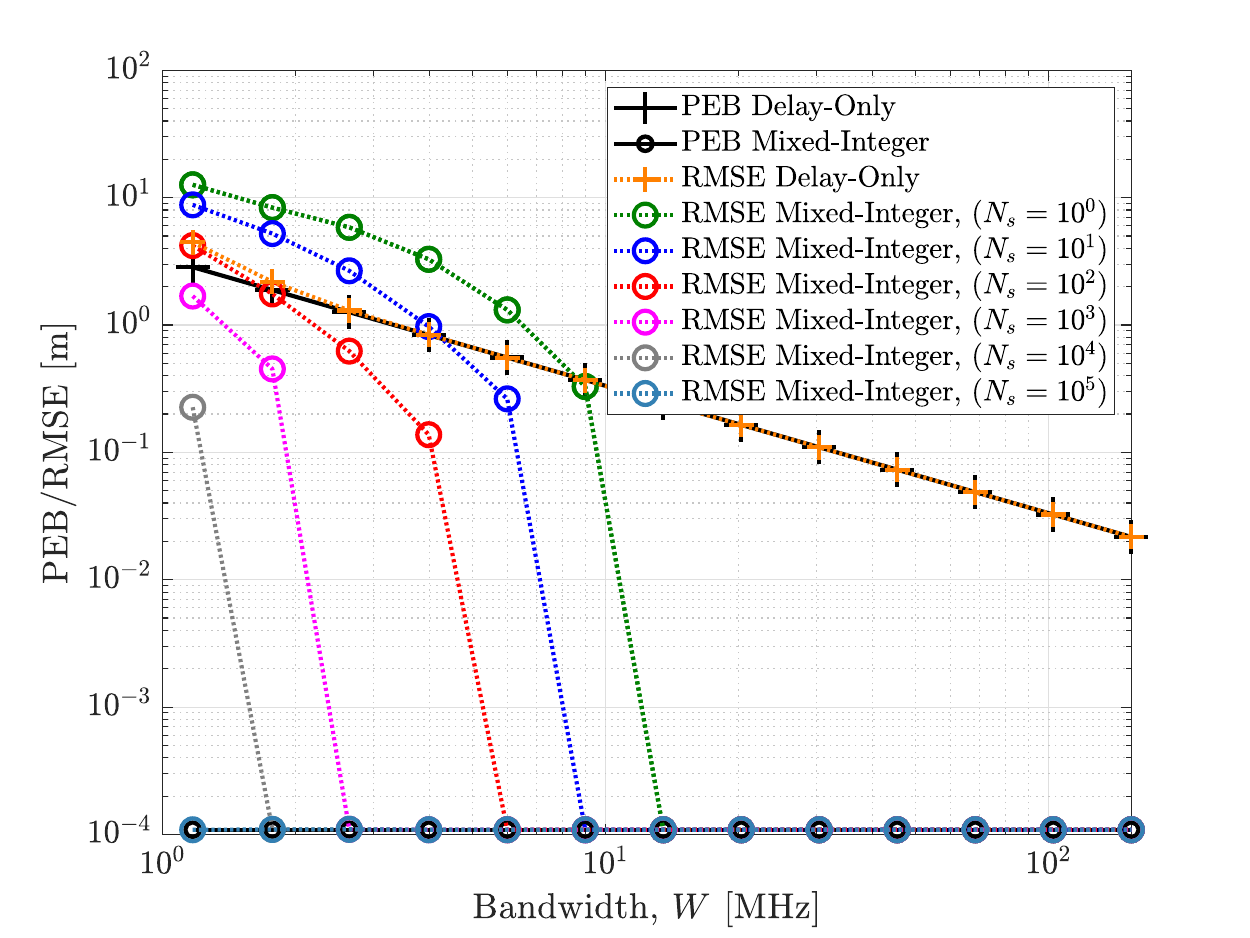}
		\vspace{-0.0mm}
        \caption{The random-search refinement results for different bandwidths in a dual-band FR1$\&$FR3 case. The $x$-axis shows the bandwidth for each band.   }
        \vspace{-4mm}
		\label{fig:searchBW}
\end{figure}

\subsubsection{Number of Second Stage Iterations}
%As illustrated in Algorithm \ref{alg:algorithm1}, the second stage of the proposed CPP algorithm is iterative, aiming to refine the position estimate. 
We next study the impact of the number of iterations, $N_{\rm iter}$, by evaluating the RMSE as $N_{\rm iter}$ increases in an example scenario with a minimal bandwidth of $13.7 \, \rm MHz$ ($456$ subcarriers). As shown in Fig.~\ref{fig:niter}, all cases except the single-band FR1 converge to their corresponding MICRB already by the second iteration. Nevertheless, the iterative algorithm also appears promising for the single-band case, as convergence starts from the third iteration. To ensure fairness, the total transmit power per BS is kept identical across all scenarios. In the following results, we thus set $N_{\rm iter}=2$, while noting that a more general stopping criterion could also be developed and applied.
\begin{figure}[!t]
		\centering
		\vspace{-1mm}
        \includegraphics[width=8.2cm,trim={0cm 0.2cm 1cm 0.5cm}]{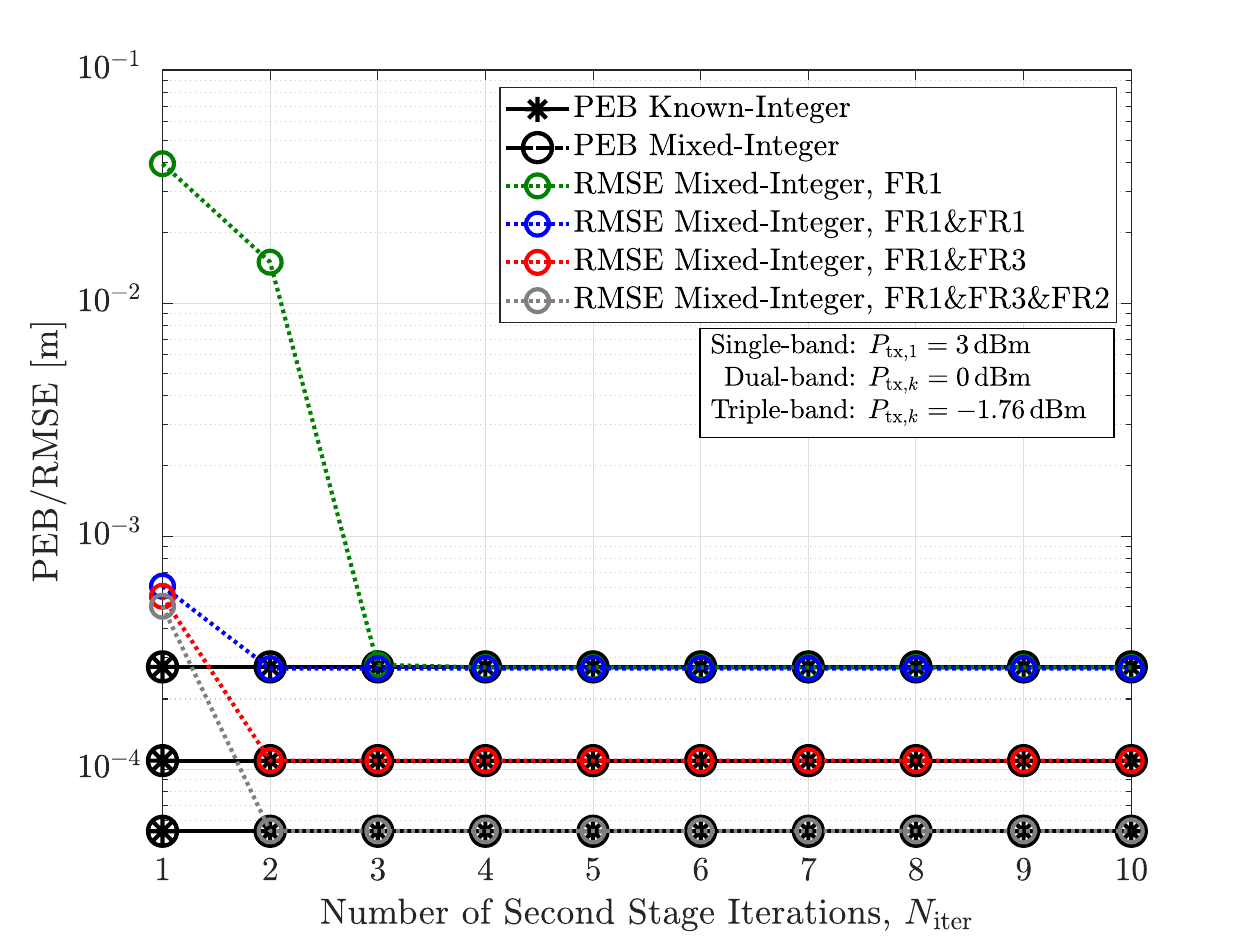}
		\vspace{-0.0mm}
        \caption{Impact of the number of second stage iterations on estimator performance. Single-band FR1 corresponds to $f_c = 3.5 \, {\rm GHz}$, FR1$\&$FR1 to $f_c = \{3.5, 3.6\} \, {\rm GHz}$, FR1$\&$FR3 to $f_c = \{3.5, 12\} \, {\rm GHz}$, and the triple-band FR1$\&$FR3$\&$FR2 case to $f_c = \{3.5, 12, 28.1\} \, {\rm GHz}$. }
        \vspace{-4mm}
		\label{fig:niter}
\end{figure}

\subsubsection{Power Budget Analysis}
We next consider default practical bandwidths for each frequency range, i.e., $20 \, {\rm MHz} $ for FR1 and $100 \, {\rm MHz} $ for FR2. Under these settings, the accuracy of the proposed estimator closely aligns with the MICRB across different per-band transmit power levels. Note that the carrier phase and delay error variances are related to the TX power via \eqref{eq:error_time}-\eqref{eq:SNR}. As shown in Fig.~\ref{fig:RMSE-Pt}, the first RMSE curve corresponds to single-band FR1 transmission. While it follows the MICRB, compared to the dual-band cases, its convergence to the asymptotic RMSE occurs more slowly (starting from $0 \, \text{dBm}$), and also the asymptotic RMSE is higher. In dual-band configurations, the estimator reaches lower asymptotic RMSE values more quickly. Specifically, FR1$\&$FR1 achieves faster convergence (starting from $-9 \, \text{dBm}$), whereas FR1$\&$FR2,  results in a lower asymptotic RMSE, at the cost of a slightly slower convergence rate (starting from $-6 \, \text{dBm}$), due to the higher carrier frequency of FR2. %For all results shown, $N_{s} = 1$ is used in the second stage. 

\begin{figure}[!t]
		\centering
		\vspace{-1mm}
        \includegraphics[width=8.2cm,trim={0cm 0.2cm 1cm 0.5cm}]{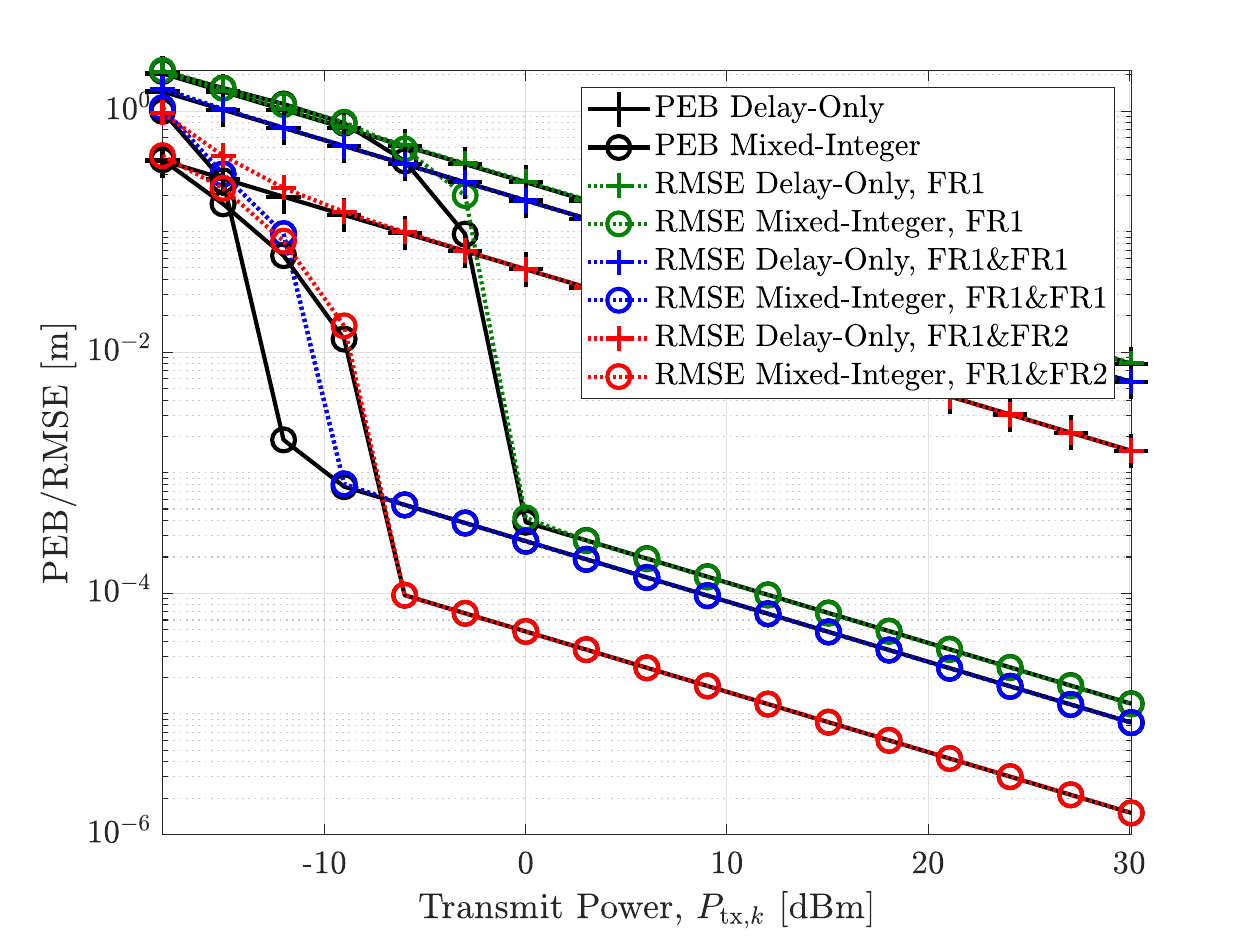}
		\vspace{-0.0mm}
        \caption{RMSE versus transmit power in different CA cases. Single-band FR1 corresponds to $f_c = 3.5 \, {\rm GHz} $ , FR1$\&$FR1 to $f_c = \{3.5, 3.6\}  \, {\rm GHz} $, and  FR1$\&$FR2 to $f_c = \{3.5, 28 \} \, {\rm GHz} $. }
        \vspace{-5mm}
		\label{fig:RMSE-Pt}
\end{figure}

\subsubsection{Nonuniform BS-Band Assignment}
\textls[-6]{We now consider an observation model where the phase offset $\varphi_{\rm{ue}}$ is band-independent. Five different example BS-band assignment patterns are analyzed and considered, illustrated in Table~\ref{ta:BS2band}, to demonstrate that the proposed bounds remain valid under such situations too, and that the estimators converge to the bounds successfully. First, we examine an extreme case where each BS is assigned to a distinct band in the range $f_c = 3.5{:}4{:}23.5 \,\rm{GHz}$, referred to as the {6b–1BS/b} pattern. Note that this scenario is feasible if and only if the phase offset is band-independent. We also consider the {6b-2BS/b} pattern for the same set of bands, where each band is assigned to two BSs; since there are $M=6$ BSs, each BS operates in a dual-band mode. Next, we evaluate a {3b-2BS/b} pattern, where three bands with $f_c = 3.5{:}4{:}11.5\,\rm{GHz}$ are each shared by two BSs (each BS operating within a single band). We also consider the fully uniform {3b-6BS/b} assignment pattern of this triple-band case, where each BS operates in the same three bands. A fully uniform assignment case of a single-band scenario (denoted as {1b-6BS/b}), in which all BSs operate on the same band at $f_c = 3.5\,\rm{GHz}$ is also considered. 
%A summary of the described patterns is provided in Table~\ref{ta:BS2band}.  %In all configurations, the system provides six delay and six phase measurements.
The bandwidth of each band is around $100 \, \rm{MHz}$ (3168 subcarriers). To ensure a fair comparison, the $x$-axis represents the total per-BS transmit power (rather than per-band transmit power $P_{{\rm tx},k}$), so that all BS-band assignments are evaluated under the same total power budget.}
%As shown in Fig.~\ref{fig:nonuniform}, performance improves as more BSs share the same band. This is because the nuisance term $\varphi_{\rm{ue}} \lambda_k $ becomes more coherent across measurements. In the single-band case, this term is identical in all phase observations and can be treated as a constant offset, which simplifies estimation and reduces uncertainty. On the other hand, the RMSE of the first stage (RMSE delay-only) output is the same in all scenarios, since the delay measurements are independent of the band.
\begin{table*}[t]
\centering
\footnotesize
%\scriptsize
\caption{\textsc{Considered BS-band assignment patterns}}
\label{ta:BS2band}
\renewcommand{\arraystretch}{1.0}
\setlength{\tabcolsep}{5pt}
\begin{tabular}{llcccccc}
\hline
\textbf{Assignment} & \textbf{Type of allocation} & \textbf{BS1} & \textbf{BS2} & \textbf{BS3} & \textbf{BS4} & \textbf{BS5} & \textbf{BS6}  \\
\hline
6b–1BS/b & 6-band disjoint & $3.5$ &$ 7.5$ & $11.5$ & $15.5$ & $19.5$ &$ 23.5$ \\

6b–2BS/b & 6-band nonuniform &$ 3.5, 7.5$ & $7.5, 11.5$ & $11.5, 15.5$ & $15.5, 19.5$ & $19.5, 23.5$ & $23.5, 27.5$ \\

3b–2BS/b & triple-band nonuniform &$ 3.5$ & $3.5 $& $7.5$ & $7.7$ & $11.5$ &$ 11.5$  \\

3b–6BS/b & triple-band uniform & $3.5, 7.5, 11.5$ & $3.5, 7.5, 11.5$ & $3.5, 7.5, 11.5 $& $3.5,7.5, 11.5$ & $3.5, 7.5, 11.5$ & $3.5, 7.5, 11.5$  \\

1b–6BS/b & single-band uniform &$ 3.5$ & $3.5 $& $3.5$ & $3.5$ & $3.5$ & $3.5$  \\
\hline
\multicolumn{8}{l}{\footnotesize 6b–1BS/b means 6 bands with 1 BS per band; 6b–2BS/b means 6 bands with 2 BSs per band, etc. All frequencies are in $\rm GHz$} \\
\hline
\end{tabular}
\end{table*}

As shown in Fig.~\ref{fig:nonuniform}, the delay-only RMSE is identical for all patterns, stemming essentially from the constant sum-power. The {6b-2BS/b} configuration outperforms {6b-1BS/b}, {3b-2BS/b}, and {1b-6BS/b}, highlighting the benefit of multi-band operation. In other words, {6b-1BS/b} and {3b-2BS/b} correspond to nonuniform single-band cases, while {6b-2BS/b} represents a nonuniform dual-band case. The {6b-2BS/b} pattern also achieves a lower converged RMSE than the fully uniform {3b-6BS/b} case because some BSs operate at higher carrier frequencies in {6b-2BS/b}. However, the {3b-6BS/b} configuration converges faster than {6b-2BS/b}, mainly due to its lower average carrier frequency and the fact that it involves triple-band operation. 

Finally, comparing {1b-6BS/b}, {3b-2BS/b}, and {6b-1BS/b} assignments, we can observe that the uniform case {1b-6BS/b} converges faster than the other ones, again due to its lower carrier frequency. The same reasoning explains the faster convergence of {3b-2BS/b} compared to {6b-1BS/b}. It should be noted that the multi-band advantage is not observed in the {3b-2BS/b} and {6b-1BS/b} configurations, since multi-band benefits arise only when an individual BS has access to multi-band measurements, allowing for the interrelated ambiguities of the same BS to be resolved jointly \footnote{\textls[-8]{Ambiguities of different BSs are already related through the geometry, and assigning each BS to a different band does not introduce additional interrelations.}}. 
% This is because the nuisance term $\varphi_{\rm{ue}} \lambda_k $ becomes more coherent across measurements. In the single-band case, this term is identical in all phase observations and can be treated as a constant offset, which simplifies estimation and reduces uncertainty. On the other hand, the RMSE of the first stage (RMSE delay-only) output is the same in all scenarios, since the delay measurements are independent of the band.

\begin{figure}[!t]
		\centering
		\vspace{-1mm}
        \includegraphics[width=8.2cm,trim={0cm 0.2cm 1cm 0.5cm}]{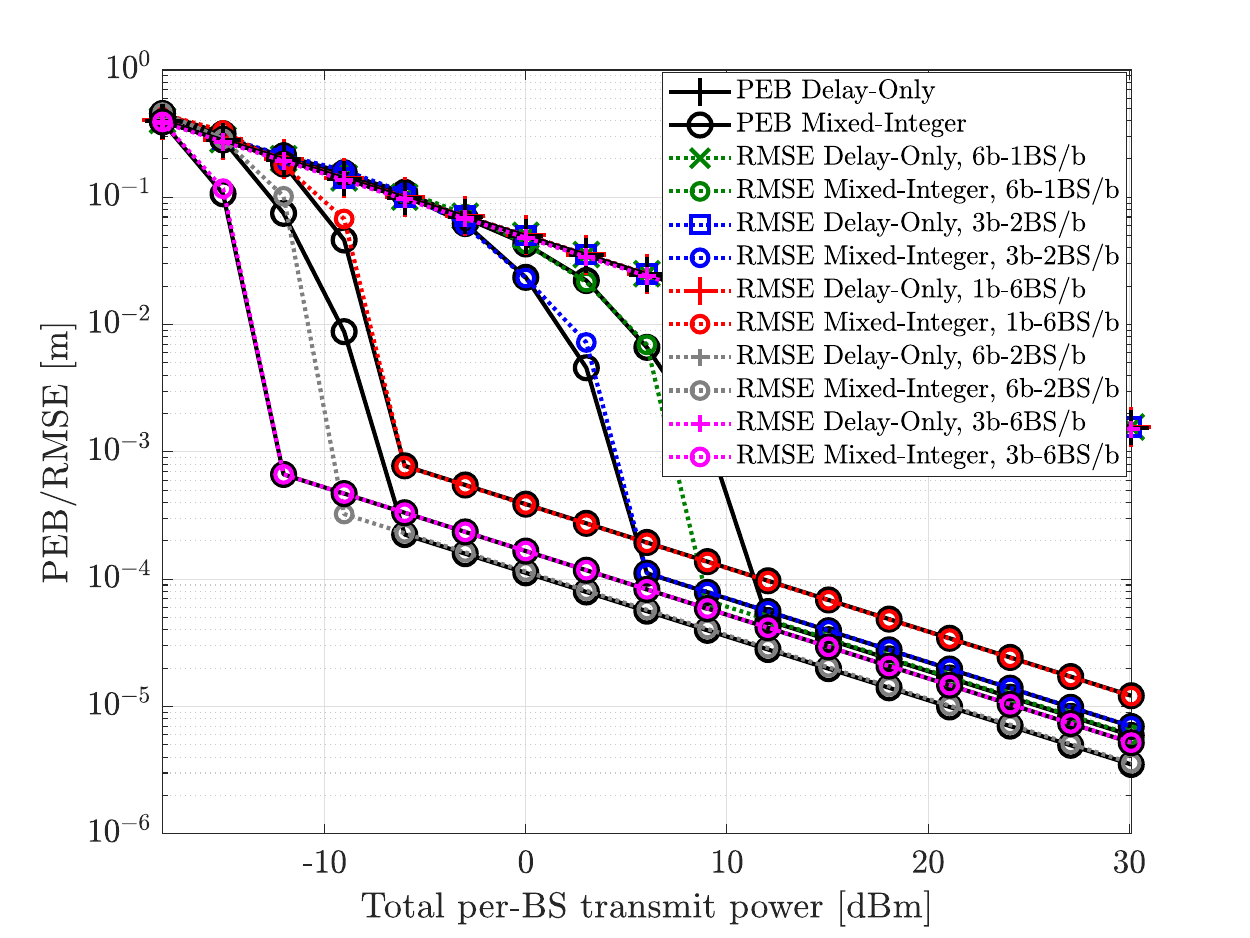}
		\vspace{-0.0mm}
        \caption{RMSE versus total per-BS transmit power for nonuniform BS-Band assignment. All the curves corresponds to the band-independent phase offset scenarios, while the details of the assignments are as shown in Table~\ref{ta:BS2band}. }
        \vspace{-4mm}
		\label{fig:nonuniform}
\end{figure}

\begin{figure*}[!t]
\centering
\vspace{-3mm}
\subfloat[Network clock uncertainties]{%
  \includegraphics[width=5.9cm,trim={0cm 0.2cm 1cm 0.5cm}]{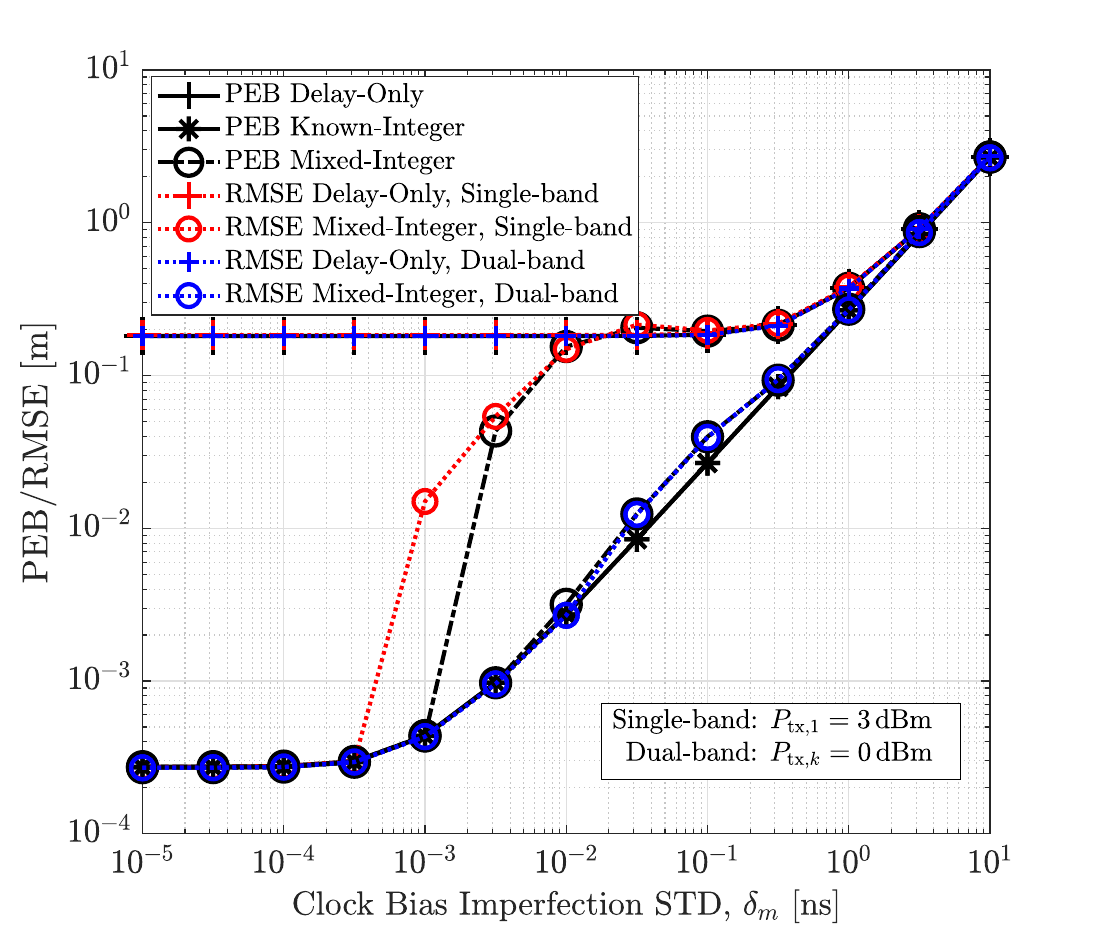}
  \label{fig:NW_clock}}
%\hspace{0.02\textwidth}
\subfloat[Multipath delay difference]{%
  \includegraphics[width=5.9cm,trim={0cm 0.2cm 1cm 0.5cm}]{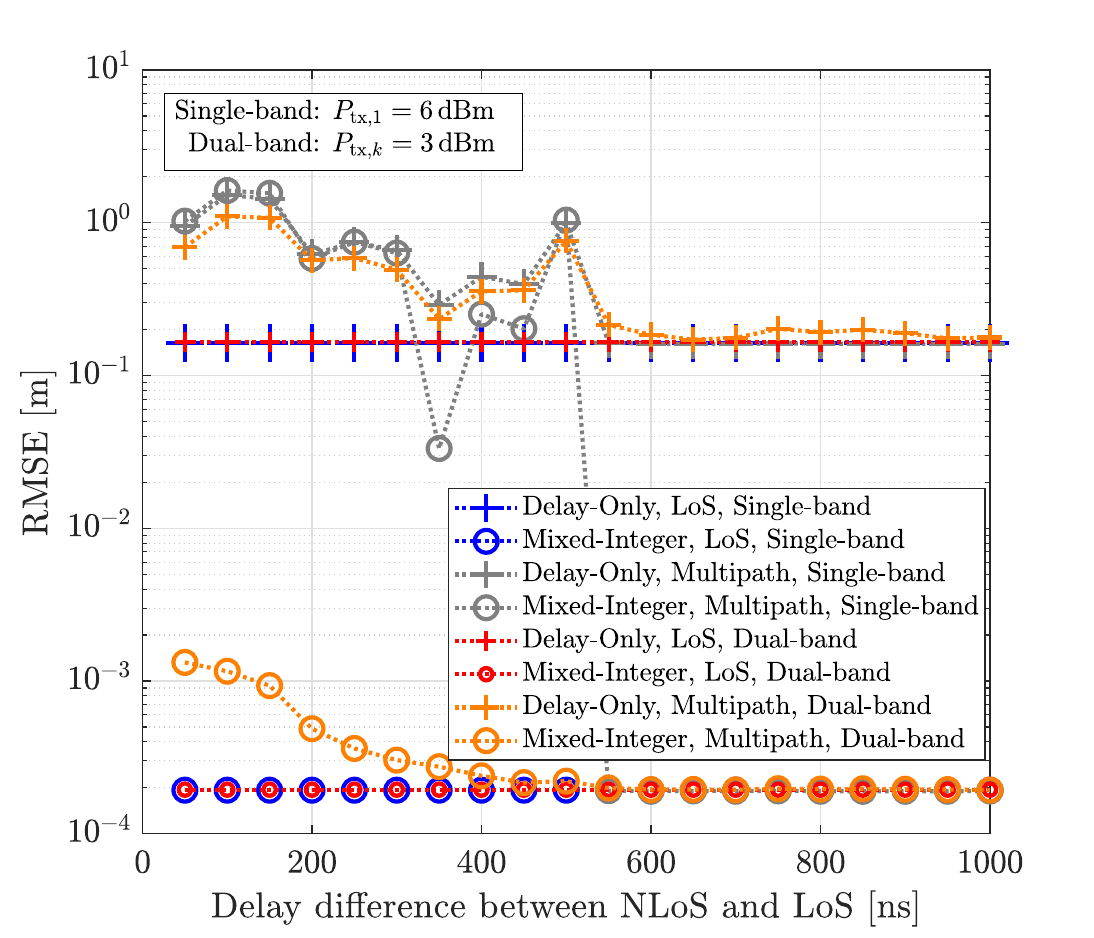}
  \label{fig:Multipath1}}
%\hspace{0.02\textwidth}
\subfloat[Multipath power ratio]{%
  \includegraphics[width=5.9cm,trim={0cm 0.2cm 1cm 0.5cm}]{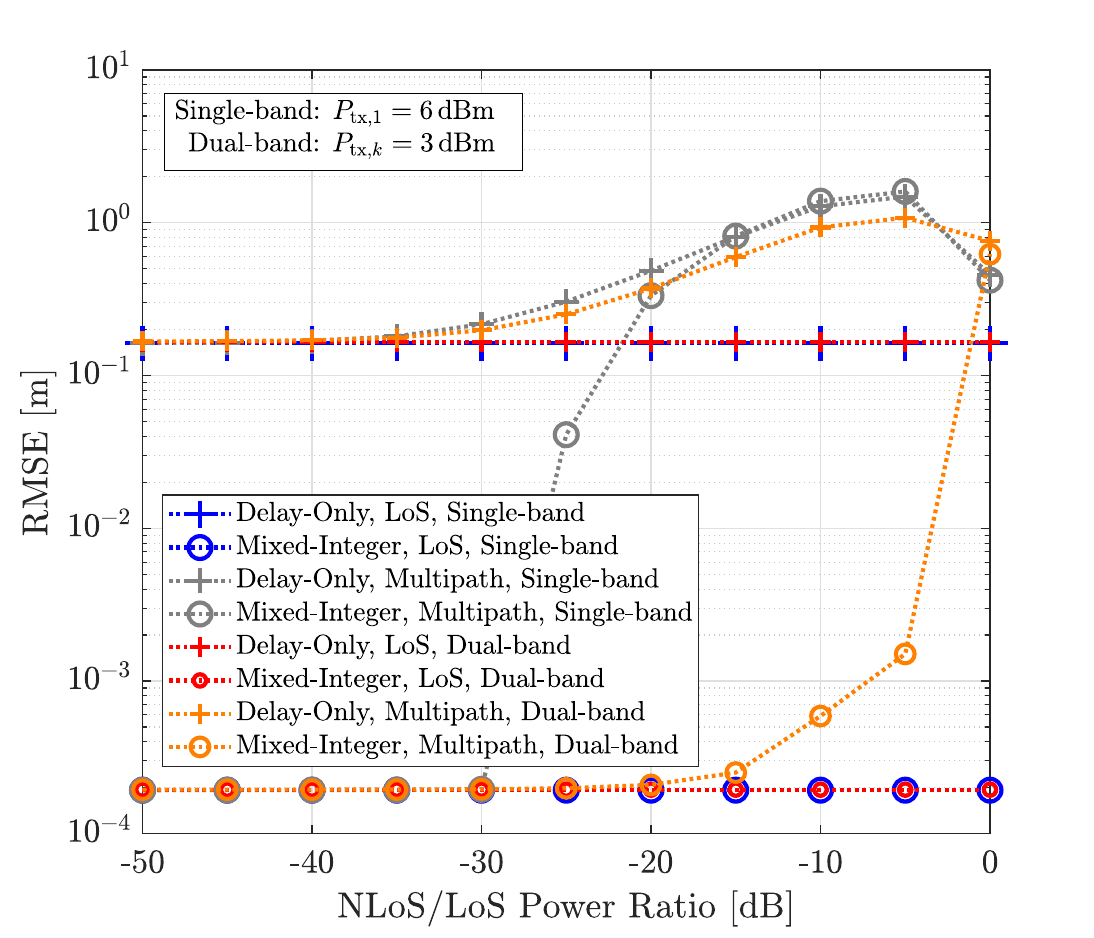}
  \label{fig:Multipath2}}
\caption{RMSE performance versus different imperfections. Single-band case corresponds to $f_c = 3.5 \, \rm{GHz}$, and the dual-band case uses  $f_c = \{3.5, 3.6\} \, \rm{GHz}$. For fair comparison, the single-band case is evaluated with doubled $P_{\rm tx}$.}
\label{fig:sensitivity_results}
\vspace{-4mm}
\end{figure*}

\vspace{-5mm}
\subsection{Estimator Sensitivity Analysis}
%common descriptions here
Finally, we analyze the sensitivity of the proposed estimator to imperfections caused by network-side clock uncertainties and multipath propagation, while comparing the dual-band and single-band configurations. To ensure a fair comparison, the total transmit power is again kept identical in both cases.
\subsubsection{Sensitivity to Network Clock Imperfections}
We model clock bias imperfections as normally distributed random offsets\cite{9566601}, common across all bands but independent across BSs. To account for this in both the bound derivation and the estimator design, the error covariance matrices $\bm \Sigma_\tau, \bm \Sigma_\vartheta$, and $\bm{\Sigma}_{\rm ch}$ are updated according to subsection  \ref{clock-imp}.  In Fig.~\ref{fig:NW_clock}, we assess and show the estimator and bound performance for different values of standard deviation (STD) of the clock uncertainty, $\delta_m$, which is assumed to be identical across all BSs.  

\textls[-3]{As depicted, CPP results are considerably more sensitive to clock imperfections than TDoA-based localization since the clock jitter affects both delay and phase measurements. Since the variance of phase errors is much smaller than that of delay errors, the impact on CPP performance becomes more pronounced. However, compared to the single-band case, the multi-band mechanism improves the robustness of CPP against clock uncertainties, even though the total per-BS transmit power is the same. Notably, in the depicted scenario, when the synchronization error is less than 100 {picoseconds}, the multi-band CPP method remains robust and achieves an acceptable RMSE performance. Such synchronization accuracy is feasible, as demonstrated in \cite{9566601}. %Finally, since the ${\rm PEB}_{\rm known}$, ${\rm PEB}_{\rm delay}$, and delay-only RMSE curves coincide for both dual-band and single-band cases, we conclude that the robustness gain of multi-band CPP originates from its ability to enhance integer ambiguity resolution.
Finally, since the ${\rm PEB}_{\rm delay}$ and delay-only RMSE curves coincide for both dual-band and single-band cases, and the ${\rm PEB}_{\rm known}$ curves are also nearly identical (indistinguishable in the plot) for both, whereas the ${\rm PEB}_{\rm known}$ and mixed-integer RMSE curves are clearly different between the two cases, we conclude that the robustness gain of multi-band CPP against the network synchronization imperfections originates from its ability to enhance integer ambiguity resolution.}

% \begin{figure}[!t]
% 		\centering
% 		\vspace{-1mm}
%         \includegraphics[width=8.2cm,trim={0cm 0.2cm 1cm 0.5cm}]{sensitivity_3.eps}
% 		\vspace{-1.5mm}
%         \caption{RMSE versus varying STDs of clock bias imperfection. The single-band case corresponds to $f_c = 3.5 \, \rm{GHz}$, and the dual-band case uses  $f_c = \{3.5, 3.6\} \, \rm{GHz}$. For fair comparison, the single-band case is evaluated with doubled $P_{\rm tx}$.}
%         \vspace{1mm}
% 		\label{fig:clock-bias-sensitivity}
% \end{figure}

\subsubsection{Sensitivity to Multi-path Propagation}
 We consider a dual-path scenario consisting of a LoS and one non-line-of-sight (NLoS) component.  As in the previous cases, the measurements are restricted to a single OFDM symbol.  The dominant paths, along with their corresponding delay and carrier phase measurements, are detected and estimated using the Hankel-based ESPRIT algorithm described in \cite{8957110,911454}. The path with the lowest delay is then selected as the LoS, and positioning is performed based on the estimated parameters of this path. We analyze the performance of the dual-band and single-band CPP under different delay differences between the LoS and NLoS paths as well as different NLoS-to-LoS power ratios. The LoS-only scenarios are included as benchmarks.

In Fig.~\ref{fig:Multipath1}, the sensitivity of the proposed estimator to NLoS interference is illustrated as a function of the delay difference between the LoS and NLoS paths. The NLoS-to-LoS power ratio is $-6 \, \rm dB$ in this example. As the delay difference increases, the two paths become more distinguishable, and in both the single-band and dual-band cases the mixed-integer RMSE of the multi-path scenario converges to that of the LoS case. % In contrast, the delay-only RMSE does not necessarily approach the LoS benchmark, indicating its weaker capability to mitigate multi-path effects.
A key observation is the superior robustness of the dual-band estimator across the entire range of delay differences. In particular, for delay differences below $500 \, \rm ns$, the single-band case shows limited robustness, while the dual-band case maintains stable performance. Although for very short delays (less than $200 \, \rm ns$) the dual-band mixed-integer RMSE exceeds the LoS benchmark, it still outperforms the delay-only estimator and achieves an accuracy around $1$ millimeter.
% \begin{figure}[!t]
% 		\centering
% 		\vspace{-1mm}
%         \includegraphics[width=8.2cm,trim={0cm 0.2cm 1cm 0.5cm}]{RMSE_multipath_delay5.eps}
% 		\vspace{-1.5mm}
%         \caption{RMSE versus varying levels of delay difference between LoS and NLoS paths. The single-band case corresponds to $f_c = 3.5 \, \rm{GHz}$, and the dual-band case uses  $f_c = \{3.5, 3.6\} \, \rm{GHz}$. For fair comparison, the single-band case is evaluated with doubled $P_{\rm tx}$.}
%         \vspace{1mm}
% 		\label{fig:multipath-sensitivity}
% \end{figure}

%\subsubsection{Multi-path Sensitivity Analysis with Varying ISR}
\textls[-3]{In Fig.~\ref{fig:Multipath2}, the delay difference between the NLoS and LoS paths is fixed at $100 \, \rm ns$ (corresponding to $30 \, \rm m$), while the NLoS-to-LoS power ratio is varied. 
The curves show that the single-band CPP estimator exhibits poor robustness and fails to outperform the delay-only benchmark for NLoS/LoS power ratios greater than $-20 \, \rm dB$. In contrast, the dual-band estimator %demonstrates strong robustness up to a value close to $-5 \, \rm dB$ because it 
consistently outperforms the delay-only RMSE of the pure LoS scenario up to NLoS/LoS power ratio of $-5 \, \rm dB$. For higher  NLoS/LoS power ratios, the NLoS component acts as strong interference. Since the path delays are close to each other, resolving the LoS and NLoS paths becomes increasingly difficult.} 
% \begin{figure}[!t]
% 		\centering
% 		\vspace{-1mm}
%         \includegraphics[width=8.2cm,trim={0cm 0.2cm 1cm 0.5cm}]{RMSE_multipath_ISR4.eps}
% 		\vspace{-1.5mm}
%         \caption{RMSE versus varying levels of ISR. The single-band case corresponds to $f_c = 3.5 \, \rm{GHz}$, and the dual-band case uses  $f_c = \{3.5, 3.6\} \, \rm{GHz}$. For fair comparison, the single-band case is evaluated with doubled $P_{\rm tx}$.}
%         \vspace{1mm}
% 		\label{fig:multipath-sensitivity-ISR}
% \end{figure}
}

% \begin{figure*}[!t]
% \centering
% %\vspace{-3mm}
% \subfloat[Network clock uncertainties]{%
%   %\includegraphics[width=0.3\textwidth]{Figures/CE_resultG_latexver2.pdf}%
%   \includegraphics[width=5.9cm,trim={0cm 0.2cm 1cm 0.5cm}]{sensitivity_3.eps}
%   \label{fig:NW_clock}}
% %\hspace{0.02\textwidth}
% \subfloat[Multipath delay difference]{%
%   %\includegraphics[width=0.3\textwidth]{Figures/CE_resultH_latexver2.pdf}%
%   \includegraphics[width=5.9cm,trim={0cm 0.2cm 1cm 0.5cm}]{RMSE_multipath_delay5.eps}
%   \label{fig:Multipath1}}
% %\hspace{0.02\textwidth}
% \subfloat[Multipath power ratio]{%
%   %\includegraphics[width=0.3\textwidth]{Figures/CE_resultF_latexver2.pdf}%
%   \includegraphics[width=5.9cm,trim={0cm 0.2cm 1cm 0.5cm}]{RMSE_multipath_ISR4.eps}
%   \label{fig:Multipath2}}
% \caption{RMSE versus different imperfections. Single-band case corresponds to $f_c = 3.5 \, \rm{GHz}$, and the dual-band case uses  $f_c = \{3.5, 3.6\} \, \rm{GHz}$. For fair comparison, the single-band case is evaluated with doubled $P_{\rm tx}$.}
% \label{fig:sensitivity_results}
% \vspace{-1mm}
% \end{figure*}

\vspace{-5mm}
\section{Conclusion}
\label{section:sec7}
\textls[-4]{This work addressed multi-band cellular carrier-phase positioning, integrating both intra-band and inter-band carrier aggregation to enhance positioning accuracy.  We introduced and derived the fundamental mixed-integer Cramér-Rao bound, specifically tailored for multi-band positioning. As critical performance benchmarks, we also derived the lower bounds for delay-only and known-integer ambiguity cases. 
{\color{black}{We then proposed a practical two-stage CPP estimator. By incorporating a search-based optimization in the second stage, we demonstrated that the estimator remains efficient even in low-bandwidth regimes}}. The offered results highlight the advantages of multi-band positioning measurements, demonstrating their ability to improve localization accuracy over single-band approaches -- particularly, in terms of resolving integer ambiguities and thereby reducing positioning errors. Additionally, we showed the impact of different carrier frequencies, carrier bandwidths, and the number of base stations, offering valuable design insights for next-generation positioning systems. Notably, we showed that aggregating just two carriers can already significantly improve integer ambiguity resolution {\color{black}while also offering improved robustness against system imperfections imposed by network-side clock uncertainties and multi-path propagation. Finally, we extended the bound derivations and CPP estimator study to scenarios where different base-stations may use partially or totally disjoint frequencies.
Future work will focus on developing reduced-complexity yet efficient CPP estimators, while also extending the CPP study to dynamic scenarios and tracking of moving UEs.
}}

\vspace{-7mm}
\appendices 
\section{Multi-band Relaxed Integer Ambiguity Bound}
\label{apdx a}
We consider $\bm z_{\rm rlx}$ as a real-valued parameter vector in the observation model (\ref{eq:diff_rlx}). Accordingly, the unknown parameter vector is defined as $\bm \eta = [\tilde{\bm{s}}^\top,{\bm{z}_{\rm rlx}}^\top]^\top\in \mathbb{R}^{(KM+{N_d}+1)\times 1}$, and the corresponding FIM matrix can be expressed as
\begin{equation}
\label{j_rlx}
\begin{split}
&{\bm{J}}_{\text{\rm rlx}}(\bm{\eta}) = \\
& \begin{bmatrix}
    {\bm{\tilde{U}}} \bm{J} {\bm{\tilde{U}}}^\top & c {\bm{\tilde{U}}}   \text{diag}(\bm{{J}}) & %\frac{1}{2\pi} ((\bm{\tilde{\Lambda}} \cdot {\bm{\tilde{U}}} )  \text{diag}(\bm{J}_{\vartheta}))
    { {\bm{\tilde{U}}}   \bm{J}_{\vartheta}{\bm{{\Lambda}}}^\top}
    \\
    c ({\bm{\tilde{U}}}   \text{diag}(\bm{{J}}))^\top & {\rm tr}({\bm{J}}) c^2 & {c (\bm{{\Lambda}}   \text{diag}(\bm{J}_{\vartheta}))^\top}  \\
   %\frac{1}{2\pi} ((\bm{\tilde{\Lambda}} \cdot {\bm{\tilde{U}}} )  \text{diag}(\bm{J}_{\vartheta}))^\top 
   %
   { \bm{{\Lambda}}({\bm{\tilde{U}}}   \bm{J}_{\vartheta})^\top}
   &  {c \bm{{\Lambda}}   \text{diag}(\bm{J}_{\vartheta})}&  {{\bm{{\Lambda}}}\bm{J}_{\vartheta}   %{\bm{\tilde{\Lambda}}}{\bm{\tilde{\Lambda}}}^\top
   {\bm{{\Lambda}}}^\top}
\end{bmatrix}
\end{split}
\end{equation}
where  $\bm{J}=\bm{J}_{\vartheta}+\bm{J}_{\tau}$, and $\bm{J}_{\vartheta}=\bm\Sigma_{\vartheta} ^{\rm -1}$. Note that ${\bm{\Lambda}}$, ${\bm{\tilde{U}}}$, and $\bm{J}_{\tau}$ are already defined in (\ref{eq:obs_model}), (\ref{j_delay}). By applying the well-known Schur complement %\cite{zhang2006schur} 
to the FIM, it can be shown that the lower bound for the error covariance of $\bm z_{\rm rlx} \in \mathbb{R}^{KM\times 1}$
is then of the form 
\begin{equation}
\begin{split}
    \bm{\Sigma}_{\rm{rlx}} & =  \bm{\Lambda}^{-1}\bm{\Sigma}_{\vartheta}\bm{\Lambda}^{-1} + \\ &  \bm{\Lambda}^{-1} 
    \begin{bmatrix}
        \bm{\tilde{U}}^\top & c\bm 1_{KM\times 1}
    \end{bmatrix}    
    \bm{J}_{\text{delay}}^{-1}(\tilde{\bm{s}}) 
    \begin{bmatrix}
        \bm{\tilde{U}} \\ c\bm 1_{1\times KM}
    \end{bmatrix}
    \bm{\Lambda}^{-1}.
\end{split}
\label{eqn:float-bound}
\end{equation}
Furthermore, for the observation model \eqref{eq:diff_rlx}, it can be shown that the error covariance bound on $\tilde{\bm s}$ is the same as the delay-only case in (\ref{j_delay}).

\section{Multi-band Known Integer Ambiguity Bound}
\label{apdx b}
When $\bm{z}_{\rm  d}$ is known in the observation model \eqref{eq:diff}, the FIM of ${\bm{s}}$ is given by
\begin{equation}
\label{j_known}
\begin{split}
&{\bm{J}}_{\text{known}}(\bm{s}) = \\
& \begin{bmatrix}
    {\bm{\tilde{U}}} \bm{J} {\bm{\tilde{U}}}^\top & c {\bm{\tilde{U}}}   \text{diag}(\bm{{J}}) & %\frac{1}{2\pi} ((\bm{\tilde{\Lambda}} \cdot {\bm{\tilde{U}}} )  \text{diag}(\bm{J}_{\vartheta}))
    {%\frac{1}{2\pi}
    {\bm{\tilde{U}}}   \bm{J}_{\vartheta}\hat{\bm{{\Lambda}}}^\top}
    \\
    c ({\bm{\tilde{U}}}   \text{diag}(\bm{{J}}))^\top & {\rm tr}({\bm{J}}) c^2 & {%\frac{c}{2\pi}
    c(\bm{\tilde{\Lambda}}   \text{diag}(\bm{J}_{\vartheta}))^\top}  \\
   %\frac{1}{2\pi} ((\bm{\tilde{\Lambda}} \cdot {\bm{\tilde{U}}} )  \text{diag}(\bm{J}_{\vartheta}))^\top 
   %
   {%\frac{1}{2\pi} 
   \bm{\tilde{\Lambda}}({\bm{\tilde{U}}}   \bm{J}_{\vartheta})^\top}
   &  {%\frac{c}{2\pi}
   c\bm{\tilde{\Lambda}}   \text{diag}(\bm{J}_{\vartheta})}& %\frac{1}{\left(2\pi\right)^2}
   {{\bm{\tilde{\Lambda}}}\bm{J}_{\vartheta}   %{\bm{\tilde{\Lambda}}}{\bm{\tilde{\Lambda}}}^\top
   {\bm{\tilde{\Lambda}}}^\top}
\end{bmatrix}
\end{split}
\end{equation}
where ${\bm{\tilde{\Lambda}}}\mathcal{}~=~ \text{diag}([\lambda_1,\cdots,\lambda_K]^\top)\otimes{\bm 1}_{ {1\times M}}\in {\mathbb{R}}^{ K \times {KM}}$.
Taking the inverse of ${{\bm{J}}_{\text{known}}(\bm{s})}$, the error covariance bound on the position reads as ${{\bm{\Sigma}}_{\text{known}}(\bm{x}_{{\rm ue}})}~= ~[{{\bm{J}}_{\text{known}}(\bm{s})}]^{-1}_{1:{N_d},1:{N_d}} $. % We can treat this bound as an \emph{optimistic} lower bound in our analysis, where the integer ambiguities are correctly resolved. 
%\balance

\bibliographystyle{IEEEtran}
\bibliography{IEEEabrv,mybibfile}

\end{document}

%% file: acronyms.tex
\acrodef{TOA}[ToA]{time-of-arrival}